\documentclass[a4paper,fleqn,usenatbib]{mnras}

\usepackage{newtxtext,newtxmath}
\usepackage[T1]{fontenc}
\usepackage{ae,aecompl}

\usepackage{graphicx}	% Including figure files
\usepackage{amsmath}	% Advanced maths commands
\usepackage{amssymb}	% Extra maths symbols

\usepackage{hyperref}
\usepackage[utf8]{inputenc}
\usepackage{mathrsfs}
\usepackage{bm}
\usepackage{hyperref}
\usepackage[dvipsnames]{xcolor}

\graphicspath{{./images/}}

\newcommand{\model}[2]{L#1\_r1}

\newcommand{\modelpns}[1]{L#1\_r2S}
\newcommand{\modelfe}[1]{L#1\_r2M}
\newcommand{\modelweak}[1]{L#1\_r2W}

\newcommand{\be}{\begin{equation}}	
\newcommand{\ee}{\end{equation}}	
\newcommand{\tx}{\text}		
\newcommand{\refig}[1]{Fig.~\ref{#1}}
\newcommand{\refeq}[1]{Eq.~\ref{#1}}

\title[Non-dipolar magnetic fields in CCSN]{The impact of non-dipolar magnetic fields in core-collapse supernov\ae}

\author[M. Bugli et al.]{
M.~Bugli,$^{1}$\thanks{E-mail: matteo.bugli@cea.fr}
J.~Guilet,$^{1}$
M.~Obergaulinger, $^{2,3}$
P.~Cerd\'a-Dur\'an $^2$
and M.A. Aloy$^{2}$
\\
$^{1}$Laboratoire AIM, CEA/DRF-CNRS-Universit\'e Paris Diderot, IRFU/D\'epartement d'Astrophysique, CEA-Saclay F-91191, France\\
$^{2}$Departamento de Astronomía y Astrof\'isica, Universitat de Val\`encia, Dr. Moliner 50, 46100, Burjassot, Spain\\
$^{3}$Institut fur Kernphysik, Theoriezentrum, Schlossgartenstr. 2, D-64289 Darmstadt, Germany
}

\date{Accepted XXX. Received YYY; in original form ZZZ}
\pubyear{2019}

\begin{document}
\label{firstpage}
\pagerange{\pageref{firstpage}--\pageref{lastpage}}
\maketitle

\begin{abstract}
The magnetic field is believed to play an important role in at least some core-collapse supernov\ae\ if its magnitude reaches $10^{15}\,\rm{G}$,  which is a typical value for a magnetar.
In the presence of fast rotation, such a strong magnetic field can drive powerful jet-like explosions if it has the large-scale coherence of a dipole. 
The topology of the magnetic field is, however, probably much more complex with strong multipolar and small-scale components and the consequences for the explosion are so far unclear. 
We investigate the effects of the magnetic field topology on the dynamics of core-collapse supernov\ae\ and the properties of the forming proto-neutron star (PNS) by comparing pre-collapse fields of different multipolar orders and radial profiles. 
Using axisymmetric special relativistic MHD simulations and a two-moment neutrino transport, we find that higher multipolar magnetic configurations lead to generally less energetic explosions, slower expanding shocks and less collimated outflows. 
Models with a low order multipolar configuration tend to produce more oblate PNS, surrounded in some cases by a rotationally supported toroidal structure of neutron-rich material. 
Moreover, magnetic fields which are distributed on smaller angular scales produce more massive and faster rotating central PNS, suggesting that higher order multipolar configurations tend to decrease the efficiency of the magnetorotational launching mechanism. 
Even if our dipolar models systematically display a far more efficient extraction of the rotational energy of the PNS, fields distributed on smaller angular scales are still capable of powering magnetorotational explosions and shape the evolution of the central compact object.
\end{abstract}


%%%%%%%%%%%%%%%%%%%%%%%%%%%%%%%%%%%%%%%%%%%%%%%%%%

%%%%%%%%%%%%%%%%% BODY OF PAPER %%%%%%%%%%%%%%%%%%

\section{Introduction}
\label{sec:intro}
Core-collapse supernov\ae\ (CCSN), which originate from the gravitational collapse of a massive progenitor star once its iron-core becomes gravitationally unstable, are amongst the most energetic astrophysical phenomena.  
While the vast majority of CCSN are believed to be driven by the so-called neutrino-heating mechanism \citep{bethe1985,janka2012,burrows2013}, there are some very energetic sources which cannot be explained by the same engine dynamics. 
Hypernovae \citep{iwamoto1998,soderberg2006,drout2011}, for instance, are powerful supernova explosions which show kinetic energies in the ejecta up to ten times larger than regular CCSN (i.e. exceeding $10^{52}\,\rm{erg}$). 
Another example are superluminous supernov\ae \citep{nicholl2013,greiner2015}, which instead present integrated luminosities exceeding the value of $10^{49}\,\rm{erg}$ reached by ordinary supernov\ae by two orders of magnitude. 
To address this excess of luminosity, popular models invoke either strong shocks due to interaction with the circumstellar medium \citep{smith2014,inserra2017} or an energy injection by a central engine at late times \citep{kasen2010,inserra2013}.

A possible solution for these two classes of extreme events is the inclusion of strong large-scale magnetic fields, which can directly couple the still forming proto-neutron star (PNS) to its surroundings.
This coupling enables the extraction of rotational energy through various processes, such as magnetic braking or the build-up of magnetic pressure gradients through winding by differential rotation. 
Fast rotation and strong magnetic fields are the fundamental ingredients at the basis of the millisecond magnetar model \citep{usov1992,thompson1994,metzger2011}, which identifies the central engine responsible for the emission of long gamma-ray bursts (LGRB) with a fast spinning and strongly magnetized PNS. 
The model has been also used to explain anomalously energetic supernov\ae\ and late-time x-ray emission from GRBs \citep{zhang2001,gompertz2014,metzger2018}.

A number of numerical studies have been conducted in the past two decades, showing that magnetic fields can have a crucial role in the explosion dynamics during the collapse of a highly magnetised progenitor \citep{meier1976,bisnovatyi-kogan1976,mueller1979,akiyama2003,kotake2004,thompson2005,obergaulinger2006a,obergaulinger2006b,burrows2007,dessart2007,obergaulinger2009,endeve2010,guilet2011,endeve2012,winteler2012,obergaulinger2014a,mosta2014,mosta2015,obergaulinger2017a,mosta2018,obergaulinger2019}. 
Strong enough magnetic fields can energise the explosion following the collapse of a rotating massive progenitor and considerably delay the formation of a central black hole, if not completely prevent it \citep{dessart2008,obergaulinger2017a,obergaulinger2017b,aloy2019b}. 
However, it is still not clear to what extent magnetic fields can lead to the powerful explosions that should connect the formation of a magnetar to energetic relativistic outflows. 
The dimensionality of the numerical models (in many cases axisymmetric) could potentially have an impact on the development of a magnetically driven explosion, due to the role of non-axisymmetric instabilities in the collimation of the forming jet \citep[][but see \citealt{obergaulinger2019}]{mosta2014}.  

One of the most uncertain ingredients in numerical models of magnetically driven explosions is the magnetic field present at the formation of the shock, both in strength and topology. 
An initial magnetic field strength of $\sim10^{11}-10^{12}$G in the iron core can reach magnetar-like values of $\sim10^{14}-10^{15}$G at bounce solely by conservation of magnetic flux, but such strong fields are hard to justify in a stellar progenitor given current evolution models. 
For instance, it has been suggested that dynamic interactions in stellar mergers might lead to a high pre-collapse field in the progenitor, hence producing a sufficiently strong magnetisation prior to collapse \citep{langer2014,schneider2016}. 
However, the viability of such process in the amplification of the magnetic field is still not completely clear. 
It is furthermore unclear whether a strong magnetic field in the progenitor is consistent with the fast rotation needed to power extreme explosions, since observed magnetic stars are slow rotators because of magnetic braking \citep{shultz2018}.
To this day, stellar models cannot provide reliable estimates for the magnetic field in a supernova progenitor, since they necessarily need rather crude approximations to include the effects of stellar dynamos at an affordable computational cost. 
One example of such approximations is the so-called Tayler-Spruit dynamo \citep{spruit2002}, which describes the amplification of magnetic fields in non-convective layers of differentially rotating stars through the growth of an instability of the toroidal field. 
This mechanism has been used in \cite{woosley2006} to include the effects of magnetically driven mass loss and angular momentum transport into one-dimensional stellar evolution models.  
Nevertheless, it is still not clear to what extent this dynamo model can capture the fundamental dynamics occurring in stably stratified stellar layers, as the resulting field, which is mostly toroidal, can vary only along the radial direction. 
In these models, the field vanishes in convectively unstable layers because convective dynamos are not included, leading to likely unrealistic configurations of magnetised and un-magnetised layers alternating in the stellar progenitor. 

An alternative scenario for the formation of the magnetic field of magnetars is represented by dynamo processes taking place within the still forming PNS. 
At early stages after the formation of the shock, the very central part of the PNS (whose size is $\sim10\,\rm{km}$) is hydrodynamic stable and in solid body rotation. 
However, this region is surrounded by a convectively unstable layer of $\sim20\,\rm{km}$ of thickness. 
In these conditions, a convective dynamo could take place and amplify an initial weak field to magnetar-like values \citep{thompson1993,raynaud2019}. 
Beyond this layer, the entropy gradient starts to rise again, but the angular velocity profile presents a significant degree of differential rotation, which sets the conditions for the development of the magnetorotational instability (MRI), that again can lead to a significant increase of the magnetic field strength \citep[e.g.][]{balbus1998,akiyama2003,masada2007,obergaulinger2009,sawai2013,guilet2015,guilet2015a,mosta2015,rembiasz2016,rembiasz2017,reboul-salze2019}.      

Numerical simulations of magnetorotational explosions cannot describe the dynamo processes in the PNS because of a lack of numerical resolution and/or the assumption of axisymmetry. 
A strong magnetic field is therefore assumed in the initial conditions prior to collapse, which can be thought of as either the actual progenitor magnetic field or an artificial way of approximating the PNS dynamo \citep[see e.g. the discussion in][]{burrows2007}. 
Whether we consider a pre-existing magnetic field in the progenitor or one resulting from dynamics in the PNS, the topology and spatial distribution of the field lines is even less constrained than the field strength. 
A common initial setup used in magnetised core-collapse simulations employs a dipolar-like magnetic field superimposed on a hydrodynamic background (usually provided by stellar evolution models), with the field having a more or less constant strength up to a characteristic radius $r_0$ and then decaying as $\sim r^{-3}$ \citep{suwa2007,burrows2007,takiwaki2009,mosta2014,obergaulinger2014a,obergaulinger2017a,mosta2018,obergaulinger2018,obergaulinger2019,aloy2019b}. 
This magnetic field configuration is chosen for the sake of simplicity but is most probably not realistic.
Lacking fully developed three dimensional stellar evolution models where the magnetic field dynamics is fully accounted for, there is almost complete freedom to choose the topology of the initial magnetic field in the pre-supernova stage. 
If instead the magnetic field is thought to represent the action of a PNS dynamo, numerical simulations of this mechanism do not suggest the generation of a dominant aligned magnetic dipole; the resulting large-scale fields can have both strong toroidal components and non-negligible multipolar contributions \citep{raynaud2019,reboul-salze2019}, thus presenting a certain degree of complexity in their angular distribution.
While a strong toroidal component is expected to develop on dynamical time scales even for an aligned dipole (due to differential winding of poloidal field lines, the so-called $\Omega$-effect), starting with a configuration with a higher multipole order $l$ is necessary to investigate the impact of having a magnetic field at smaller angular scales.  

Despite the uncertainty in the topology of the field, there are very few examples in the literature of studies aimed at quantifying the impact of non-dipolar fields on the dynamics of CCSN.
\cite{ardeljan2005} presented a model with a quadrupolar-like field superimposed on the hydrodynamic background at the formation of the PNS. 
They reported a magnetically driven explosion where the ejecta were expelled preferentially along the equatorial direction, rather than the rotational axis. 
On the other hand, \cite{sawai2005} presented models with a similar magnetic configuration, showing that they tended to produce faster and more collimated polar outflows with respect to the dipolar case. 
The results from these two studies appear to contradict each other, although it should be considered that they employed vastly different prescriptions for the progenitor profiles, equation of state (EoS), neutrino transport and initial magnetic to rotational energy ratio. 

In this paper, we present the first study that systematically addresses the impact of non-dipolar magnetic field topologies on magnetically driven CCSN. 
We performed a series of numerical simulations that differ from each other solely by the initial magnetic field, in order to appreciate the intrinsic dependencies of the system dynamics on the field topology. 
Both the explosion and the PNS formation are significantly affected by a different distribution of the magnetic field in both radius and polar angle, even when characteristic quantities such as total magnetic energy or surface field at a given radius are kept fixed. 
After describing our numerical tools and initial models in Section \ref{sec:models}, we discuss the results of our simulations in Section \ref{sec:results}, focusing first on the onset of the explosion and then on the evolution of the central PNS. 
Finally, we present our conclusions and perspectives in Section \ref{sec:conclusions}.

%%%%%%%%%%%%%%%%%%%%%%%%%%%%%%%%%%%%%%%%%%%%%%%%%%
%%%%%%%%%%%%%%%%%%%%%%%%%%%%%%%%%%%%%%%%%%%%%%%%%%
%%%%%%%%%%%%%%%%%%%%%%%%%%%%%%%%%%%%%%%%%%%%%%%%%%
%%%%%%%%%%%%%%%%%%%%%%%%%%%%%%%%%%%%%%%%%%%%%%%%%%
%%%%%%%%%%%%%%%%%%%%%%%%%%%%%%%%%%%%%%%%%%%%%%%%%%
%%%%%%%%%%%%%%%%%%%%%%%%%%%%%%%%%%%%%%%%%%%%%%%%%%
%%%%%%%%%%%%%%%%%%%%%%%%%%%%%%%%%%%%%%%%%%%%%%%%%%
%%%%%%%%%%%%%%%%%%%%%%%%%%%%%%%%%%%%%%%%%%%%%%%%%%
%%%%%%%%%%%%%%%%%%%%%%%%%%%%%%%%%%%%%%%%%%%%%%%%%%
%%%%%%%%%%%%%%%%%%%%%%%%%%%%%%%%%%%%%%%%%%%%%%%%%%

\section{Models}
\label{sec:models}

We perform our axysimmetric simulations using the numerical code \texttt{AENUS-ALCAR} \citep{obergaulinger2014b,just2015}, which solves the equations of special-relativistic magnetohydrodynamic (SRMHD) coupled to a M1 closure scheme for the neutrino transport in an approximated Newtonian gravitational potential with relativistic corrections \citep{marek2006}. 
Since the code has been employed in the last decade for several studies on magnetised CCSN \citep{obergaulinger2006a,obergaulinger2014a,obergaulinger2018}, we will make use of an initial setup very similar to the model \emph{35OC-Rs} presented in \cite{obergaulinger2017a}, which essentially differs just for the magnetic field considered.  

%%%%%%%%%%%%%%%%%%%%%%%%%%%%%%%%%%%%%%%%%%%%%%%%%%
%%%%%%%%%%%%%%%%%%%%%%%%%%%%%%%%%%%%%%%%%%%%%%%%%%
%%%%%%%%%%%%%%%%%%%%%%%%%%%%%%%%%%%%%%%%%%%%%%%%%%

\subsection{Numerical setup}
The starting point of all our simulations is the stellar model \texttt{35OC} \citep{woosley2006}, which describes a pre-collapse massive star with a zero-age main-sequence mass of $M_\tx{ZAMS}=35M_\odot$ and whose evolution takes into account both rotation and an approximated dynamo mechanism to amplify the progenitor magnetic field. 
The transport of angular momentum mediated by the magnetic field leads to a significant mass-loss rate through stellar winds, resulting into a mass of $M_\tx{35OC}=28.1M_\odot$ at collapse. 
The progenitor has a rather large non-convective iron-core with mass $M_\tx{Fe}\sim2.1M_\odot$ and radius $R_\tx{Fe}\sim2.89\times10^8\,\rm{cm}$, which is surrounded by a convective region of about $4M_\odot$ with a flat entropy profile and a slower rotational profile than the iron-core \citep[see Fig. 1 of][for the radial profiles of density, entropy and specific angular momentum]{obergaulinger2017a}.  

To describe the nuclear matter above the density threshold $\rho_\tx{LS}=10^8\,\rm{g\ cm}^{-3}$ we employ the nuclear EOS of \cite{lattimer1991} with an incompressibility of $K=220$ MeV~(hereafter referred to as LS220). 
Although LS220 has been technically ruled out as a viable nuclear EoS due to the inconsistency of its symmetric energy parameters with constraints set by nuclear experiments \citep{tews2017} and the fact that it is non-casual at very high densities \cite[e.g.][and references therein]{aloy2019a}, we still employed it in our models in order to present a more meaningful comparison with the work of \cite{obergaulinger2017a}. 
For $\rho<\rho_\tx{LS}$ we consider the contribution of relativistic leptons, photons and baryons. 
For the latter, we assume a composition of pure $^{28}$Si for temperatures $T<0.44\,\rm{MeV}$ or pure $^{56}$Ni otherwise. 
The interaction processes between matter and neutrinos that we consider include nucleonic and nuclear scattering and absorption, inelastic scattering of electrons, electron-positron annihilation into pairs of neutrinos and anti-neutrinos and nucleon-nucleon bremsstrahlung \cite[for a complete list see][]{obergaulinger2018}. 

The numerical domain extends in the radial direction from the centre to $R_\tx{out}=1.4\times10^{10}\,\rm{cm}$ and includes the whole polar range $\theta\in[0,\pi]$ rad. 
The grid in the radial direction is logarithmically stretched such as to have equal aspect ratio of the cells down to a uniform resolution of $6\times10^4\,\rm{cm}$ in the centre of the computational domain. 
We make use of a coarsening scheme in $\theta$ in the central region for radii smaller than $\sim24\,\rm km$ in order to mitigate the CFL constraint on the time-step size. 
This particular choice of the numerical discretisation has been proven to provide sufficiently converged results in terms of the post-bounce (p.b.) dynamics \citep{obergaulinger2014a,obergaulinger2014b,obergaulinger2017a}.

%%%%%%%%%%%%%%%%%%%%%%%%%%%%%%%%%%%%%%%%%%%%%%%%%%
%%%%%%%%%%%%%%%%%%%%%%%%%%%%%%%%%%%%%%%%%%%%%%%%%%
%%%%%%%%%%%%%%%%%%%%%%%%%%%%%%%%%%%%%%%%%%%%%%%%%%

\subsection{Initial magnetic field}

A number of numerical studies employ the following prescription for the azimuthal component of the vector potential \citep{suwa2007}:
\be\label{eq:A_dipole}
A^\phi_\tx{dip}= \frac{B_0}{2}\frac{r_0^3}{r^3+r_0^3} r\sin\theta,
\ee
which, for an axisymmetric domain, results in a (purely poloidal) dipolar magnetic field whose radial component is approximately constant up to a characteristic radius $r_0$ and then decays as $\sim r^{-3}$.

To obtain a more general prescription for a multipolar initial magnetic field, let us assume axisymmetry and express $B^r$ and $B^\theta$ in terms of the vector potential as
\begin{eqnarray}
\label{eq:br-aphi}
B^r&=&(\bm{\nabla}\times\bm{A})_r=\frac{1}{r\sin\theta}\frac{\partial}{\partial\theta}(\sin\theta A^\phi),\\ 
\label{eq:bth-aphi}
B^\theta&=&(\bm{\nabla}\times\bm{A})_\theta=-\frac{1}{r}\frac{\partial}{\partial r}(rA^\phi).
\end{eqnarray}
We can express the radial and polar components of an axisymmetric and purely poloidal magnetic field with a specific multipolar component of order $l$ as
\begin{eqnarray}
B^r_l&=&\tilde{B}^r(r)P_l(\cos\theta),\\
B^\theta_l&=&\tilde{B}^\theta(r)\frac{\tx{d} P_l(\cos\theta)}{\tx{d}\theta},
\end{eqnarray}
where $P_l$ is the Legendre polynomial of order $l$ and the functions $\tilde{B}^r$ and $\tilde{B}^\theta$ contain the radial dependency of the magnetic field components.

\begin{figure}
\centering
\includegraphics[width=\columnwidth]{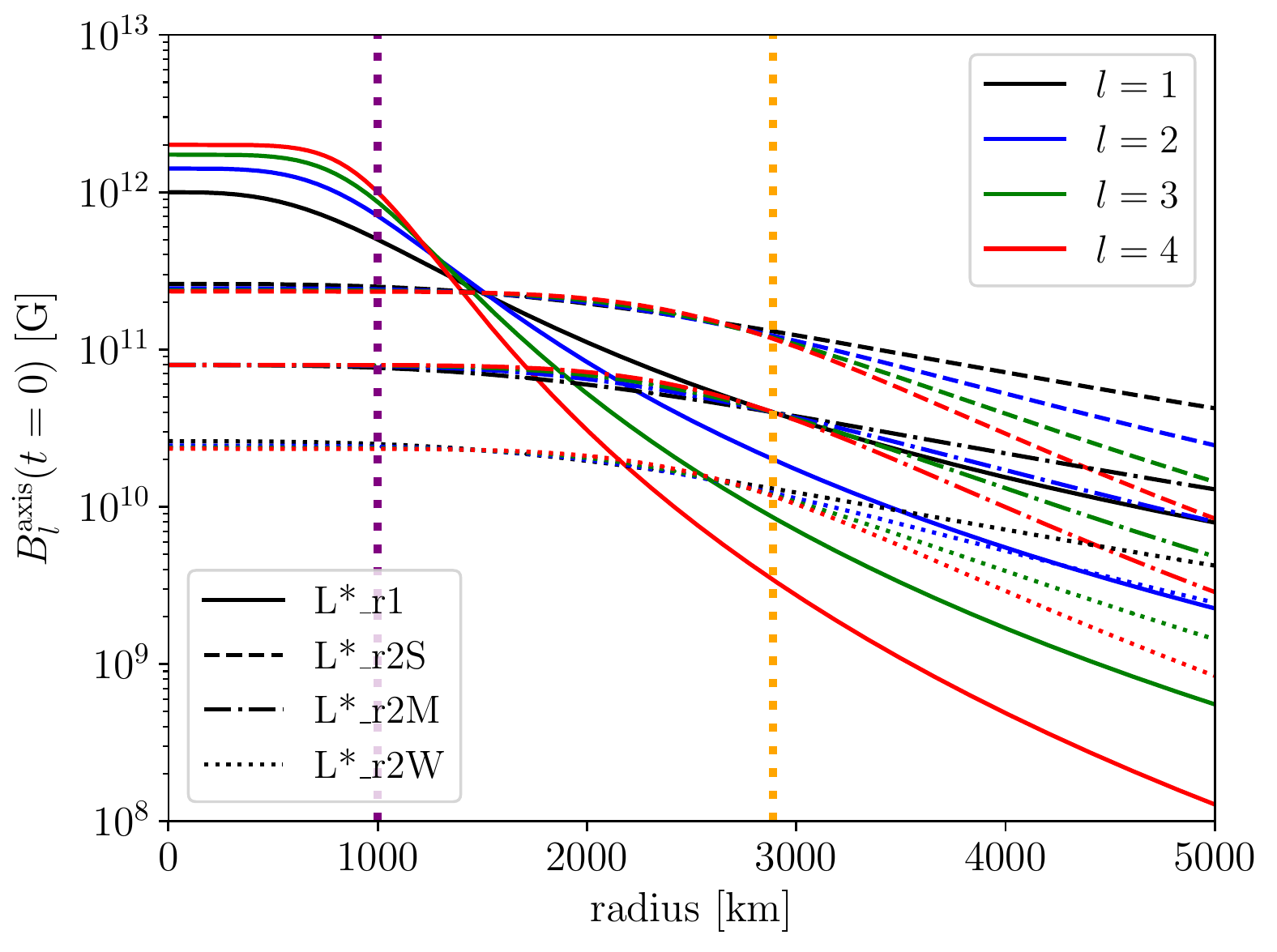}
\caption{Radial magnetic field along the rotational axis (as given by \refeq{eq:Bpoles}) for all models in the present work, at $t=0$. Black, blue, green and red lines refer respectively to $l=1,2,3,4$. 
The purple and orange dotted vertical lines represent the two characteristic radii considered in this work, i.e. $r_0=[1,2.89]\times10^3\rm\,\rm{km}$.}
\label{fig:b0t0}
\end{figure}

\begin{figure}
\centering
\includegraphics[width=\columnwidth]{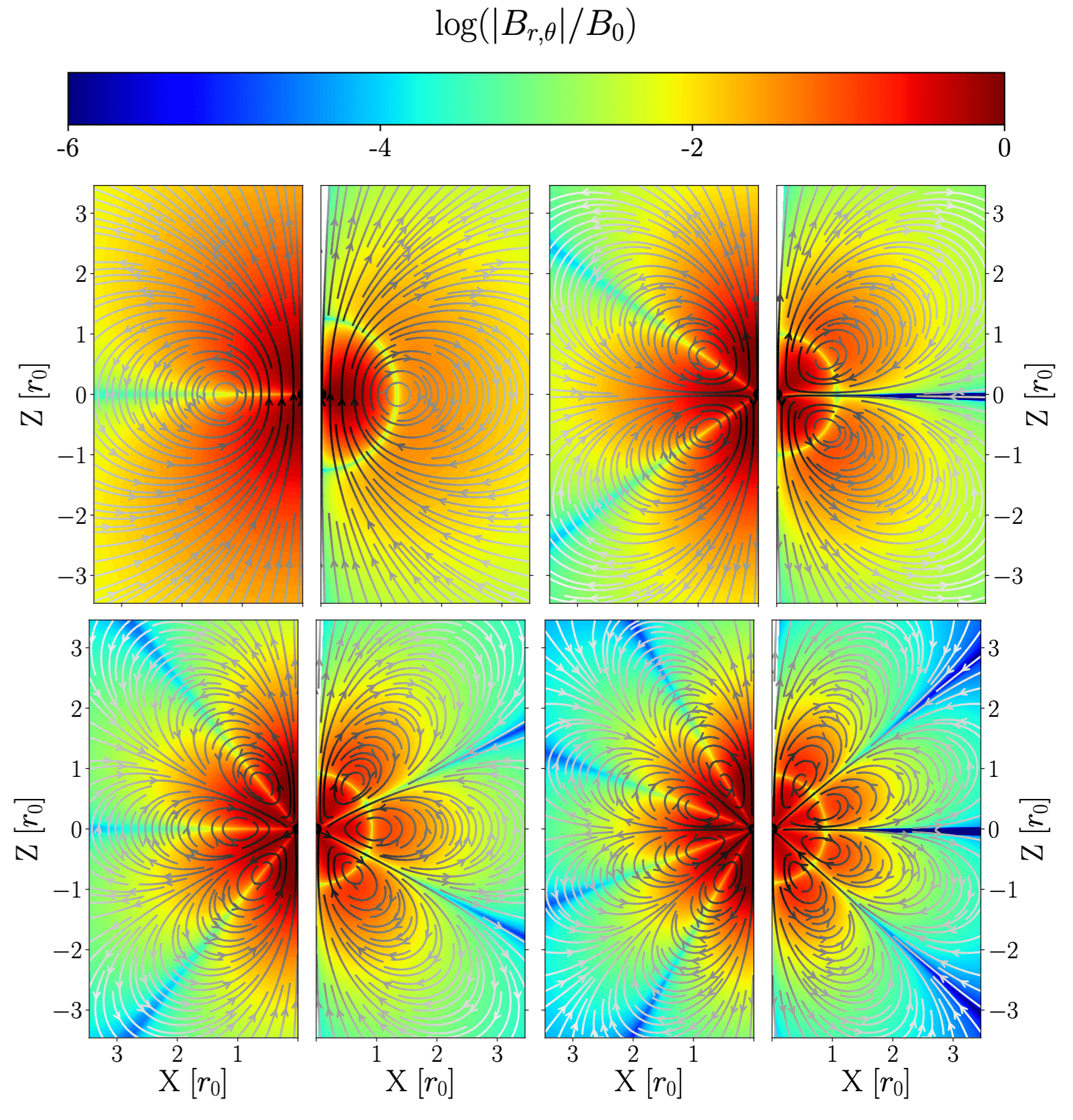}
\caption{Magnitude of the radial (left side) and polar (right side) components of the magnetic field in units of $B_0$, at $t=0$. 
Since the axes are expressed in units of $r_0$, these profiles apply to all models considered in this work. The streamlines track the initial poloidal magnetic field employed in each simulation.}
\label{fig:fields}
\end{figure}

As this study aims at assessing the importance of the magnetic field topology in the explosion mechanism and ultimately making a comparison with previous results obtained in the literature, it will be useful to make use of a general prescription for a multipolar expansion of the magnetic field which is consistent with \refeq{eq:A_dipole} for a magnetic dipole, i.e. for $l=1$. 
We then require the radial magnetic field of a given order $l$ to be
\be\label{eq:br-pl}
B^r_l=B_0\frac{r_0^{l+2}}{r^{l+2}+r_0^{l+2}}P_l(\cos\theta),
\ee
where $r_0$ is a characteristic radius and $B_0$ is a normalisation constant. For $r\gg  r_0$ the field decays as $\sim r^{-(l+2)}$, while for $r\ll r_0$ it remains constant along the radial direction. 
If we decompose the azimuthal component of the vector potential in two factors containing the dependence with the radius $r$ and the polar angle $\theta$
\be
A^\phi_l(r,\theta)=F_l(r)\tilde{A}_l(\theta),
\ee
we can use this expression with \refeq{eq:br-aphi} and \refeq{eq:br-pl} to write
\begin{eqnarray}
\tilde{A}_l(\theta)&=&\frac{1}{\sin\theta}\int_0^\theta\sin\theta' P_l(\cos\theta')d\theta',\label{eq:A_theta}\\
F_l(r)&=&rB_0\frac{r_0^{l+2}}{r^{l+2}+r_0^{l+2}}.
\end{eqnarray}
Using the following property of the Legendre polynomials
\be
(2l+1)P_l(x)=\frac{\tx{d}}{\tx{d}x}(P_{l+1}(x)-P_{l-1}(x))
\ee
to compute the integral in \refeq{eq:A_theta}, we finally obtain 
\be\label{eq:Aphi}
A^\phi_l(r,\theta)= r\frac{B_0}{(2l+1)}\frac{r_0^{l+2}}{r^{l+2}+r_0^{l+2}} \frac{P_{l-1}(\cos\theta)-P_{l+1}(\cos\theta)}{\sin\theta}.
\ee
\refig{fig:b0t0} and \refig{fig:fields} show the characteristics of the initial magnetic field configuration for all models considered in this work. 

%%%%%%%%%%%%%%%%%%%%%%%%%%%%%%%%%%%%%%%%%%%%%%%%%%
%%%%%%%%%%%%%%%%%%%%%%%%%%%%%%%%%%%%%%%%%%%%%%%%%%
%%%%%%%%%%%%%%%%%%%%%%%%%%%%%%%%%%%%%%%%%%%%%%%%%%

\subsection{Normalisation of the magnetic field}
Since we are interested in the effects of the magnetic field topology on  the explosion dynamics, we should normalise the vector potential in such a way as to keep some quantity related to the magnetic field constant for different values of the multipole expansion order $l$. 
Such a quantity could be, for instance, the total magnetic energy $E_{\tx{mag},l}^{r_0}$, i.e. the integral of the local magnetic energy density $\epsilon_{\tx{mag},l}=((B^r_l)^2+(B^\theta_l)^2)/8\pi$ within a sphere of radius $r_0$. 
From the numerical integration of the different magnetic field configurations using \refeq{eq:Aphi} we find that
\be\label{eq:E_mag}
\frac{E_{\tx{mag},l}^{r_0}}{E_{\tx{mag},1}^{r_0}}\propto l^{-1},
\ee
a scaling that derives directly from the ortho-normalisation properties of Legendre polynomials.  
Thus, by redefining  $A^\phi_l\rightarrow\sqrt{l}A^\phi_l$ we can factor out the dependence on the multipolar order.
For a given value of $r_0$ and $B_0$ and for any $l$, the magnetic energy within a sphere of radius $r_0$ is then approximately given by
\be
E_{\tx{mag},r_0}\simeq6.3\times 10^{47}\left(\frac{B_0}{10^{12}\tx{G}}\right)^2\left(\frac{r_0}{10^8\tx{cm}}\right)^3\tx{erg},
\ee
within $\sim5\%$ accuracy.
The last column of Table \ref{tab:models} reports the magnetic energy contained in a sphere enclosing a mass of $2.5M_\odot$, i.e.  $E_{\tx{mag},2.5}$, which in our model has a radius of $\sim5500 \,\rm{km}$ and represents a typical shell used in the literature to estimate the compactness of the core. 
The decreasing value of such magnetic energy with larger $l$ for models \model{\tx{*}}{12} (which share the same $B_0$ and $r_0$) is due to the different radial decay of the magnetic field beyond $r_0$. 

\begin{table}
    \caption{Parameters of the models presented in this work.}\label{tab:models}
    \centering
    \begin{tabular}{ccccccc}
        \hline
        & $N_r$ & $N_\theta$ & $l$ & $B_0$  & $r_0$  & $E_{\tx{mag},2.5}$  \\
        &   &   &   &   [$10^{10}$G]    & [$10^8$cm]    & [$10^{47}$erg]\\
        
        \hline
        \modelpns{1}	& 400 & 128 &  1  & $2.61\times10^{1}$ & 2.89 & $1.56\times10^{1}$  \\
        \modelpns{2}    & 400 & 128 &  2  & $1.73\times10^{1}$ & 2.89 & $6.27\times10^{0}$  \\
        \modelpns{3}   	& 400 & 128 &  3  & $1.37\times10^{1}$ & 2.89 & $3.84\times10^{0}$  \\
        \modelpns{4}    & 400 & 128 &  4  & $1.17\times10^{1}$ & 2.89 & $2.79\times10^{0}$  \\
        \hline
        \modelfe{1}	    & 400 & 128 &  1  & $7.97\times10^{0}$ & 2.89 & $1.46\times10^{0}$  \\
        \modelfe{2} 	& 400 & 128 &  2  & $5.64\times10^{0}$ & 2.89 & $6.66\times10^{-1}$  \\
        \modelfe{3}     & 400 & 128 &  3  & $4.60\times10^{0}$ & 2.89 & $4.33\times10^{-1}$  \\
        \modelfe{4}     & 400 & 128 &  4  & $3.99\times10^{0}$ & 2.89 & $3.25\times10^{-1}$  \\
        \hline
        \modelweak{1}	& 400 & 128 &  1  & $2.61\times10^{0}$ & 2.89 & $1.56\times10^{-1}$  \\
        \modelweak{2}  	& 400 & 128 &  2  & $1.73\times10^{0}$ & 2.89 & $6.27\times10^{-2}$  \\
        \modelweak{3} 	& 400 & 128 &  3  & $1.37\times10^{0}$ & 2.89 & $3.84\times10^{-2}$  \\
        \modelweak{4}  	& 400 & 128 &  4  & $1.17\times10^{0}$ & 2.89 & $2.79\times10^{-2}$  \\
        
        \hline
        \model{1}{12} 	& 400 & 128 &  1  & $10^{2}$          & 1 & $1.07\times10^{1}$    \\
        \model{2}{12}   & 400 & 128 &  2  & $10^{2}$          & 1 & $8.97\times10^{0}$    \\
        \model{3}{12}   & 400 & 128 &  3  & $10^{2}$          & 1 & $8.55\times10^{0}$    \\
        \model{4}{12}   & 400 & 128 &  4  & $10^{2}$          & 1 & $8.47\times10^{0}$    \\
        \hline
\end{tabular}
\end{table}

With this normalisation the strength of the magnetic field along the symmetry axis is given by
\be\label{eq:Bpoles}
B^\tx{axis}_{l}=|\bm{B}_{l}(r,\theta=0)|=\sqrt{l}\frac{B_0}{1+(r/r_0)^{l+2}}.
\ee

The parameter $r_0$, which essentially selects the radial extent of the bulk of the magnetic field in the progenitor, has been set in the literature to mainly two different values, either $3000 \,\rm{km}$ \citep{dessart2008,harikae2009} or $1000 \,\rm{km}$ \citep{obergaulinger2017a}. 
The dynamics of the collapse of the progenitor 35OC are such that the mass enclosed within $1000\,\rm{km}$ will be accreted even before bounce, while the mass initially enclosed in a sphere of $3000\,\rm{km}$ radius will accrete later than $500\,\rm{ms}$ p.b.. 
In our study we set for most of our models $r_0=2 890\,\rm{km}$, which corresponds to the initial size of the iron core $R_\tx{Fe}$: this allows us to focus on the early phases after the formation of the shock factoring out the impact of the radial decay of the magnetic field. 
On the other hand, we set $r_0=1000\,\rm{km}$ for models \model{\text{*}}{12} to include this effect and compare our results with previous studies employing such a magnetic structure.

An important quantity that could be expected to play a role in regulating the explosion dynamics is the strength of the magnetic field close to the PNS surface. 
We therefore modulate the value of $B_0$ such as to obtain the same magnetic field strength for different multipolar configurations at a given radius that should be representative of the PNS surface. 
However, it is not possible to define a priori a particular shell in the progenitor that should correspond to such a surface, since the outer layers of the star continuously accrete onto the PNS, at least, for a few seconds after bounce. 

We operated then the following choice. 
For models \modelfe{\text{*}} we set $B_0$ such as to match the magnetic field strength at the rotation axis of model \model{1}{12} at the radius $R_\mathrm{Fe}$. 
Imposing this further constraint requires of course the use of different values for $B_0$ and consequently a different content of magnetic energy amongst models \modelfe{\text{*}} (see Table \ref{tab:models}). 
We adopt a similar choice for models \modelpns{\text{*}}, but we replace the matching radius with $R_\tx{b}=1500 \,\rm{km}$.
The value of $R_\tx{b}$ defines a sphere enclosing the same mass as the PNS at the time of shock formation, such that the magnetic field at the PNS surface at shock formation is approximately the same for all models of this series. 
This choice produces a magnetic field which is larger by roughly a factor 3 with respect to models \modelfe{\text{*}}. 
 
Models with the suffix \emph{W} (standing for \emph{weak} field) relate directly to the group of models identified by \emph{S} (\emph{strong} field), but start with a magnetic field which is respectively one order of magnitude weaker than the latter. 
Models of with the suffix \emph{M} (\emph{moderate} field) possess magnetic energies in between strong and weak model series.
See Table \ref{tab:models} for a complete listing of the models we studied and the parameters that characterise them. 
 
\begin{figure}
\centering
\includegraphics[width=0.95\columnwidth]{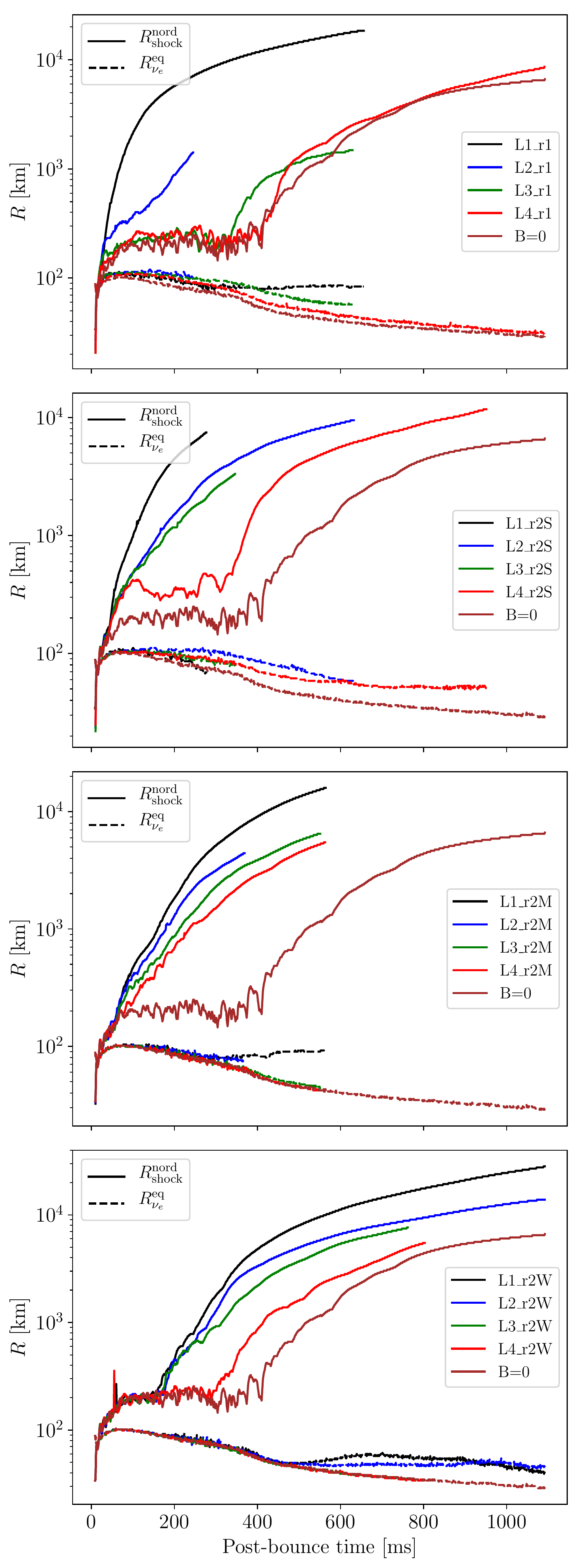}
\caption{Time evolution of the shock radius along the symmetry axis in the north hemisphere (solid lines), and of the electron neutrino-sphere along the equator (dashed lines). 
The top panel refers to different multipolar configurations with a small value for $r_0$ and a strong field (models \model{\text{*}}{12}), while the other three show (from top to bottom) models with large $r_0$ and progressively weaker magnetic field strength.}
\label{fig:radii}
\end{figure}

%%%%%%%%%%%%%%%%%%%%%%%%%%%%%%%%%%%%%%%%%%%%%%%%%%
%%%%%%%%%%%%%%%%%%%%%%%%%%%%%%%%%%%%%%%%%%%%%%%%%%
%%%%%%%%%%%%%%%%%%%%%%%%%%%%%%%%%%%%%%%%%%%%%%%%%%
%%%%%%%%%%%%%%%%%%%%%%%%%%%%%%%%%%%%%%%%%%%%%%%%%%
%%%%%%%%%%%%%%%%%%%%%%%%%%%%%%%%%%%%%%%%%%%%%%%%%%
%%%%%%%%%%%%%%%%%%%%%%%%%%%%%%%%%%%%%%%%%%%%%%%%%%
%%%%%%%%%%%%%%%%%%%%%%%%%%%%%%%%%%%%%%%%%%%%%%%%%%
%%%%%%%%%%%%%%%%%%%%%%%%%%%%%%%%%%%%%%%%%%%%%%%%%%
%%%%%%%%%%%%%%%%%%%%%%%%%%%%%%%%%%%%%%%%%%%%%%%%%%
%%%%%%%%%%%%%%%%%%%%%%%%%%%%%%%%%%%%%%%%%%%%%%%%%%

\section{Results}
\label{sec:results}

We discuss and compare next the results of our simulations, having as a goal the identification of specific features that can be directly connected to the topology of the magnetic field. 
In the following we will regard model \model{1}{12} as the prototype of a magnetorotational explosion, since a very similar configuration has been recently investigated by \cite{obergaulinger2017a} (employing the same numerical tool we used for this work) and has led to an explosion dominated by the magnetic field dynamics. 
We will also stress  differences and analogies that may arise between the results from our magnetised models and those coming from one without magnetic fields, which represents the opposite extreme of a standard delayed hydrodynamic neutrino-driven explosion. 
Although we will only focus on specific groups of models for the discussion of some quantities (e.g. \modelfe{\text{*}} or \modelpns{\text{*}} ), all trends coming from different values of $l$ shown in the current section are to be considered as occurring for any particular group model (unless otherwise specified).

%%%%%%%%%%%%%%%%%%%%%%%%%%%%%%%%%%%%%%%%%%%%%%%%%%
%%%%%%%%%%%%%%%%%%%%%%%%%%%%%%%%%%%%%%%%%%%%%%%%%%
%%%%%%%%%%%%%%%%%%%%%%%%%%%%%%%%%%%%%%%%%%%%%%%%%%

\subsection{Explosion dynamics}
All the simulations presented in this work led to a successful explosion, with some models producing prompt explosions and others presenting an initially stalling shock which starts to expand after $\sim400\,\rm{ms}$ p.b., at the latest (see \refig{fig:radii}). 
The simulations with initial dipolar field exhibit, with respect to their multipolar counterparts, a faster expansion of the shock radius and, in some cases, a shorter lived stalling phase. 
This trend applies to all values of $l$ we considered, showing that higher multipolar expansions are more prone to produce stalling shocks and hence delayed explosions. 
Unsurprisingly, for a given value of $l$, the stronger the initial magnetic field is, the faster is the onset and expansion of the shock.
A further confirmation of the importance of the field topology in establishing the dynamics of the explosion comes from a comparison of the shock expansion between models \modelfe{1} and \modelpns{2}. 
Despite having a higher initial magnetic energy, the quadrupolar configuration produces a relatively slower increase in shock radius with respect to the dipolar model with weaker magnetic field. 

\begin{figure}
\centering
\includegraphics[width=1.\columnwidth]{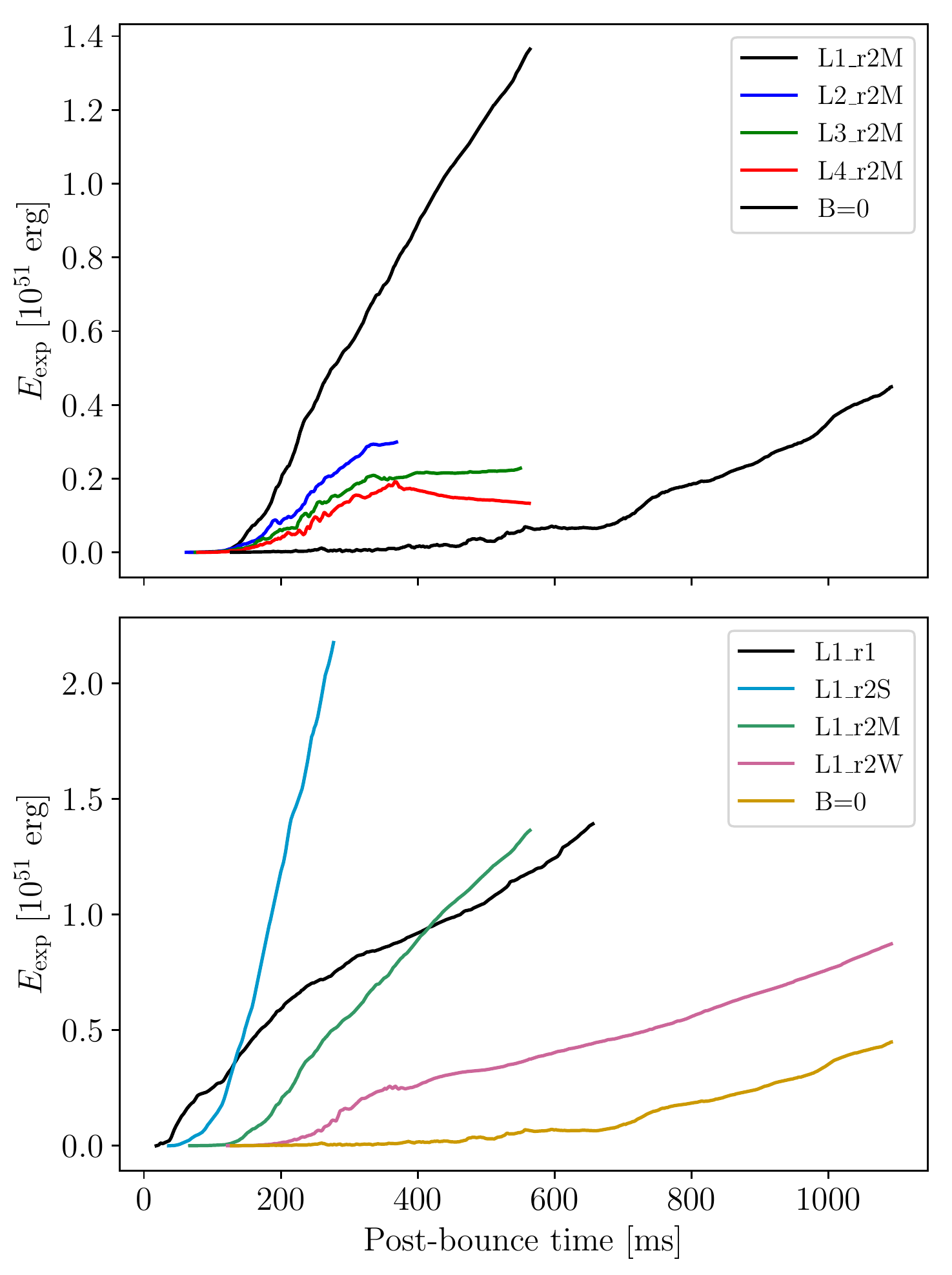}
\caption{Time evolution of the total energy of the gravitationally unbound ejecta for models \modelfe{\text{*}} (top panel) and different dipolar models (bottom panel).}
\label{fig:exp_energy}
\end{figure}

While for models \modelfe{\text{*}} the onset of the explosion occurs at the same time (third panel from the top of \refig{fig:radii}), the same cannot be said for models \model{\text{*}}{12} (top panel): the shock starts to expand almost instantaneously with a dipolar field, while it stalls essentially as long as in the hydrodynamic case for $l=4$. 
This effect is due to the relatively small value of $r_0=10^3\,\rm{km}$ and steeper radial decay of the magnetic field for higher multipolar expansions: at shock formation, all the highly magnetised material from the central regions has already collapsed into the PNS, and hence the magnetic flux being accreted from this point drops to progressively lower values for higher $l$.
This leads to model \model{4}{12} displaying a dynamic evolution very similar to the counterpart without magnetic fields.

To measure the energy of the explosion, we define the gravitationally unbound ejecta as the material with outward radial velocity and a positive value of the total energy density 
\be\label{eq:E_ej}
\epsilon_\tx{tot}=\epsilon_\tx{int}+\epsilon_\tx{kin}+\epsilon_\tx{mag}+\epsilon_\tx{grav},
\ee 
i.e. the sum of internal, kinetic, magnetic and gravitational energy densities. 
We define a proxy of the explosion energy $E_\tx{ej}$ as the volume integral of \refeq{eq:E_ej} over the region occupied by the unbound ejecta.
As for the shock radius, from \refig{fig:exp_energy} we can see that models with a higher multipolar magnetic configuration end up producing less energetic explosions. 
This trend applies no matter the specific normalisation choice adopted, suggesting that more complicated field topologies lead to a systematic weakening of the explosion. 
We can also see from the bottom panel of \refig{fig:exp_energy} that the dipolar models with a larger value of $r_0$ and sufficiently strong initial field display a much steeper rise of the explosion energy, with respect to model \model{1}{12}. 
This is likely due to the more efficient conversion of gravitational potential energy into magnetic pressure deriving from a more spatially extended magnetic field, which enables a further compression of the magnetic flux during the collapse of the highly magnetised external layers of the progenitor. 
It should be noticed that this proxy for the explosion energy does not include the gravitational binding energy of the external stellar layers excluded from our numerical box, which have a mass of about $3.8\,M_\odot$ and account for roughly $3\times10^{51}\,\rm{erg}$ in binding energy.
Accordingly, one would require longer-lasting simulations in order to investigate the value of the asymptotic explosion energies (which are clearly not achieved in the present work). 

\begin{figure}
\centering
\hspace{0.9cm}
\includegraphics[width=\columnwidth]{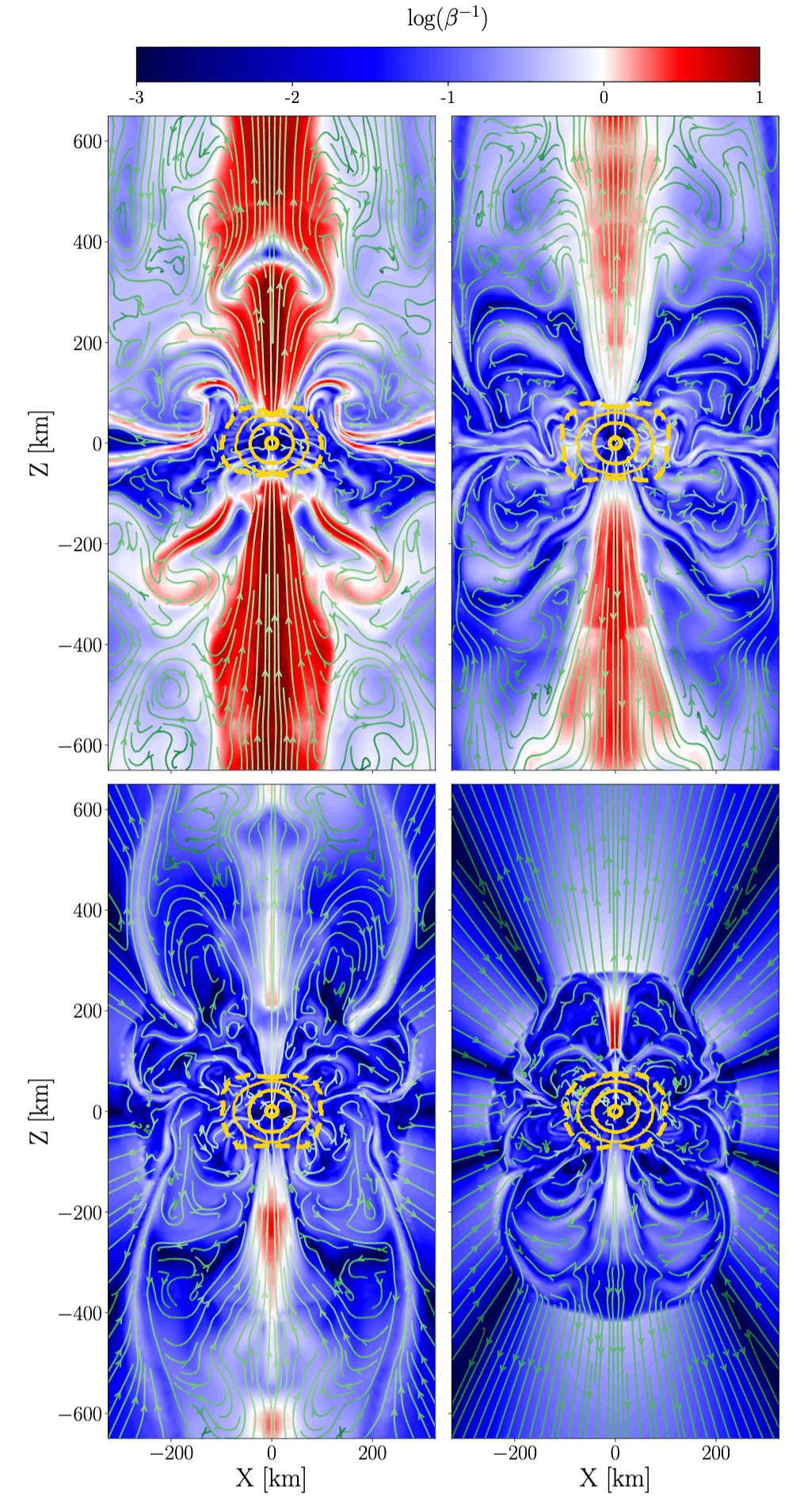}

\caption{Magnetic to thermal pressure ratio  at $t\sim158$ ms p.b., for models \modelpns{1} (top left), \modelpns{2} (top right), \modelpns{3} (bottom left) and \modelpns{4} (bottom right). The solid yellow lines represent density contours of $10^{11}$, $10^{12}$ and $10^{14}$ g cm$^{-3}$, while the dashed yellow line is the electron neutrino-sphere. The streamlines represent the poloidal component of the magnetic field. The colour bar is chosen such that white corresponds to a ratio equals to 1, thus separating domains where pressure is dominated by the thermal component (blue) and by the magnetic component (red).}
\label{fig:beta-1}
\end{figure}

The rate of expansion of the shock radius and the overall time of onset of the explosion is tightly connected to the ratio of magnetic over thermal pressure (also referred to as inverse plasma beta, i.e. $\beta^{-1}$), which is a key parameter in the dynamics of magnetorotational explosions. 
As shown in \refig{fig:beta-1}, at a given time there can be a broad range in intensity and distribution of $\beta^{-1}$ among models \modelpns{\text{*}}, which all start with relatively strong magnetic fields and accrete up to $\sim400\,\rm{ms}$ p.b. highly magnetised material (since this is roughly the time-scale required for the collapse and accretion onto the PNS of the whole iron core). 
We observed a systematic decrease in the magnetic pressure support of the outflow with increasing order of multipolar expansion in all the models we considered. 
This effect could be related to the fact that a magnetic field defined on a larger angular scale connects a larger fraction of the surroundings of the PNS to the polar region, hence leading to a more effective piling up of the magnetic pressure  which ultimately is responsible for the launching of the outflow.

\begin{figure}
\centering
\includegraphics[width=1.\columnwidth]{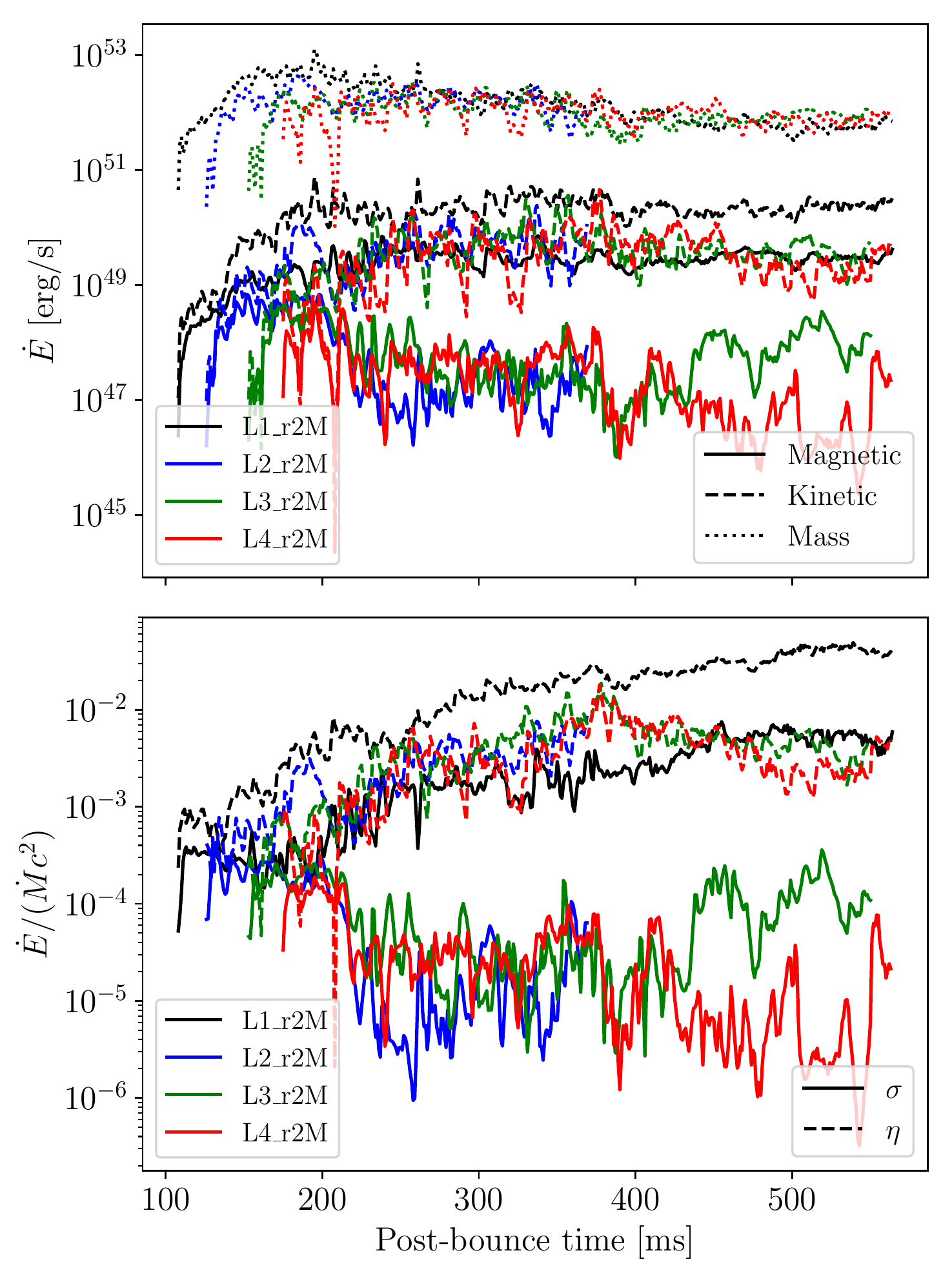}
\caption{In the top panel we show the evolution with time of the Poynting flux, kinetic energy flux and mass flux computed at a distance of $500\,\rm{km}$ from the centre and averaged over an half-opening angle of $10^\circ$. 
In the bottom panel are reported the magnetisation and baryon loading parameter as defined by \refeq{eq:sigma}.
All quantities refer to models \modelfe{\text{*}}.
}
\label{fig:sigma}
\end{figure}

We conduct a further analysis on the outflow properties by computing the magnetisation $\sigma$ and the baryon loading parameter $\eta$, defined as
\be\label{eq:sigma}
\sigma=\frac{\dot{E}_\tx{mag}}{\dot{M}c^2},\qquad \eta=\frac{\dot{E}_\tx{kin}}{\dot{M}c^2},
\ee 
where $\dot{E}_\tx{mag}$ is the Poynting flux, $\dot{E}_\tx{kin}$ is the flux of kinetic energy and $\dot{M}$ is the mass flux. 
We compute $\sigma$ and $\eta$ only in the regions occupied by the ejecta, averaging the fluxes in terms of which they are defined over a cone with half-opening of $10^\circ$.
From the top panel of \refig{fig:sigma} we see that 
for any given model the flux of kinetic energy systematically dominates over the Poynting flux in all models considered. 
Both of them are also much smaller than the corresponding mass flux, and consequently both $\sigma$ and $\eta$ are considerably smaller than unity.
This feature is to be expected, since none of the outflows produced in our models is relativistic.
 
We notice, however, a clear distinction between the dipolar case and the other multipolar configurations.
Amongst models \modelfe{\text{*}} only the $l=1$ case shows a systematic growth of both $\sigma$ and $\eta$, reaching maximum values of $\sim7\times10^{-3}$ and $\sim4\times10^{-2}$ respectively, after $\sim500\,\rm{ms}$.
For models with higher multipolar configurations, instead, the magnetisation decreases considerably in the first few 100 ms and sets at 2-3 orders of magnitude below the value reached in the dipolar case; $\eta$, on the other hand, appears to reach a maximum around $\sim10^{-2}$ at $t\sim380\,\rm{ms}$ and then decreases slightly.
These differences in the growth of $\sigma$ and $\eta$ between different multipoles can be better understood by looking at the evolution of the Poynting flux and the kinetic luminosity of the outflow (top panel of \refig{fig:sigma}). 
While the mass flux $\dot{M}$ slightly decreases at the same pace for all models in the \modelfe{\text{*}}  series, the kinetic energy and Poynting fluxes reach a rather constant value with no significant decrease only in the dipolar case.
On the other hand, for $l>1$ the decrease in $\dot{E}_\tx{mag}$ and $\dot{E}_\tx{kin}$ compensates for the lower mass flux and consequently produce a stalling behaviour in $\sigma$ and $\eta$.
We verified that the lower value of $\dot{E}_\tx{mag}$ and $\dot{E}_\tx{kin}$ for higher multipolar models does not depend substantially on our choice for the opening angle over which we performed the average, despite models with higher values of $l$ displaying narrower magnetised regions along the symmetry axis.

\begin{figure}
\centering
\hspace{0.9cm}
\includegraphics[width=\columnwidth]{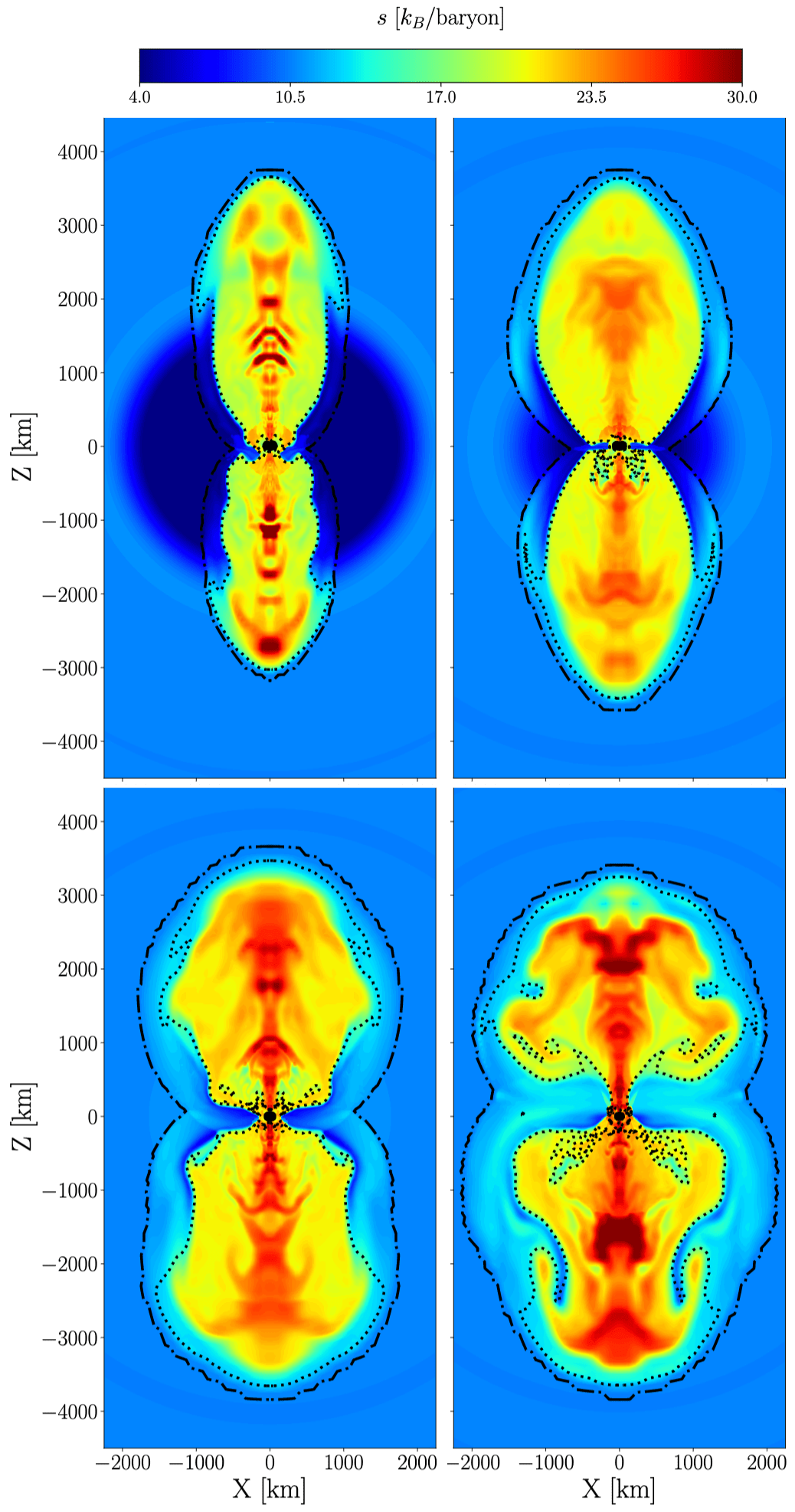}

\caption{Specific entropy for models \modelfe{1} (top left), \modelfe{2} (top right), \modelfe{3} (bottom left) and \modelfe{4} (bottom right) when the axial radius of the shock has reached $\sim3500\,\rm{km}$.}
\label{fig:entropy}
\end{figure}

\begin{figure}
\centering
\includegraphics[width=\columnwidth]{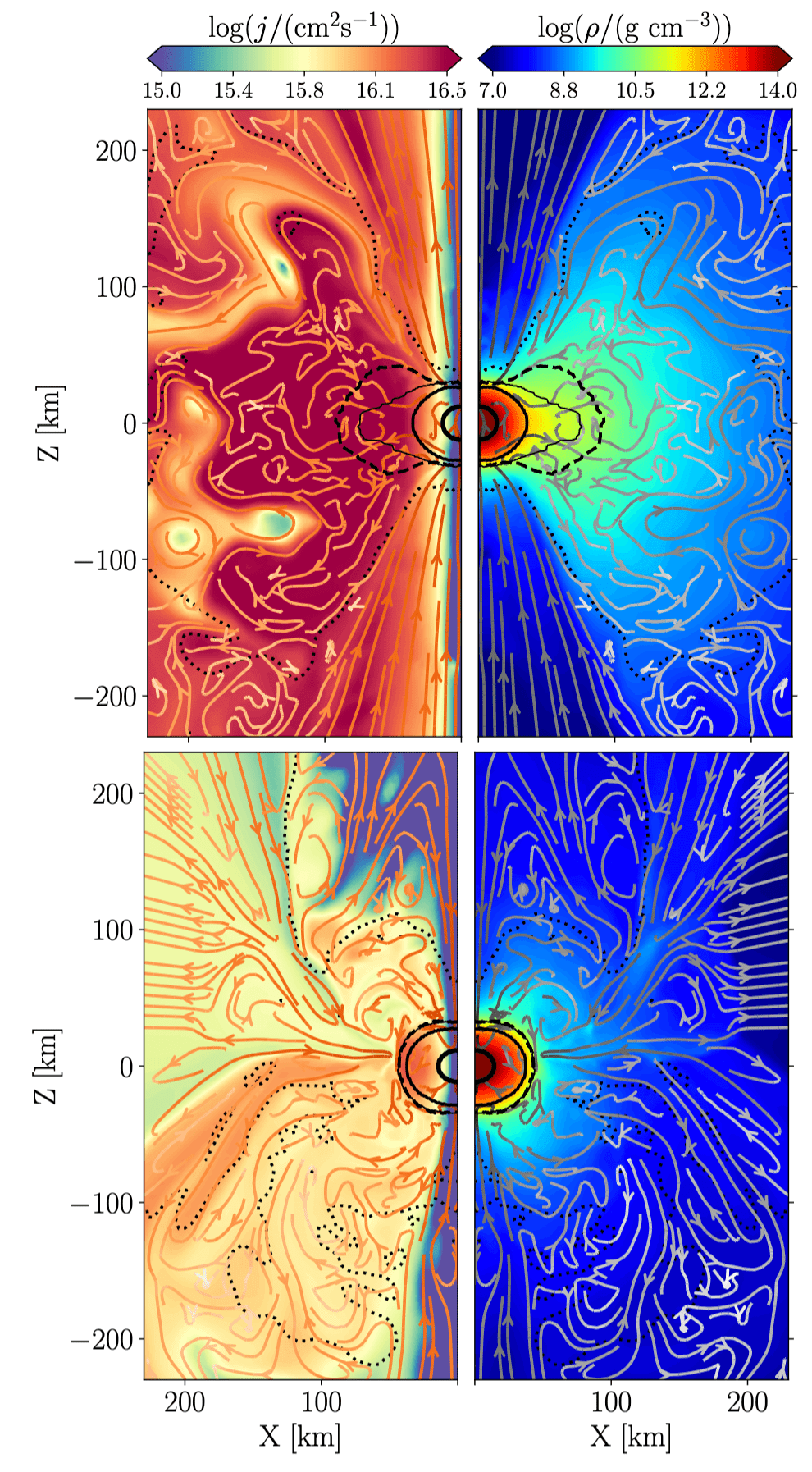}
\caption{Specific angular momentum (left column) and mass density (right) at $t\sim525$ ms p.b., for models \modelfe{1} (top row) and \modelfe{4} (bottom). 
The solid black lines represent density contours of $10^{11}$, $10^{12}$ and $10^{14}\,\rm{g\ cm}^{-3}$.
The dashed line is the electron neutrino-sphere, while the dotted lines enclose the gravitationally unbound ejecta. 
The streamlines represent the poloidal component of the magnetic field.}
\label{fig:j-rho}
\end{figure}

We shift next our attention to the degree of collimation of the polar outflows produced in our simulations. 
\refig{fig:entropy} shows the distribution of specific entropy in the ejecta produced by models \modelfe{\text{*}} when the maximum of the shock radius has reached a distance of $\sim3500\,\rm{km}$ from the center.
There is a clear correlation between the \emph{aspect ratio} of the ejecta shape (i.e. the relative size of the shock along the symmetry axis and the equator) and the order of the multipolar expansion, with on one side the dipolar case displaying a very well collimated outflow and, on the other hand, model \modelfe{4} having a much larger cocoon and complex structure in the external layers. 

This is somewhat in contrast with the findings of \cite{sawai2005}, who reported higher degrees of collimation when a quadrupolar magnetic field was employed instead of a dipolar one, but it does not confirm the results reported in \cite{ardeljan2005} either, since we do not observe an ejection of material along the equator stronger than along the polar axis.
Our results can be considered a sort of middle ground, where bipolar explosions are still to be expected from higher order multipoles, although less collimated.
It is not clear, however, if the dipolar and quadrupolar configurations considered in these studies differ exclusively by their angular distribution, or to what extent they share the same radial structure. 
If one considers, on top of this, the significant differences between \cite{ardeljan2005} and \cite{sawai2005} in the initial conditions employed (e.g. progenitor profile, radial structure and strength of the magnetic field, rotation profile), the possibility of  a more meaningful comparison between our work and the two aforementioned ones appears to be rather difficult and prone to be mostly speculative.

%%%%%%%%%%%%%%%%%%%%%%%%%%%%%%%%%%%%%%%%%%%%%%%%%%
%%%%%%%%%%%%%%%%%%%%%%%%%%%%%%%%%%%%%%%%%%%%%%%%%%
%%%%%%%%%%%%%%%%%%%%%%%%%%%%%%%%%%%%%%%%%%%%%%%%%%

\subsection{PNS formation}
\label{subsec:pns}

\begin{figure}
\centering
\includegraphics[width=\columnwidth]{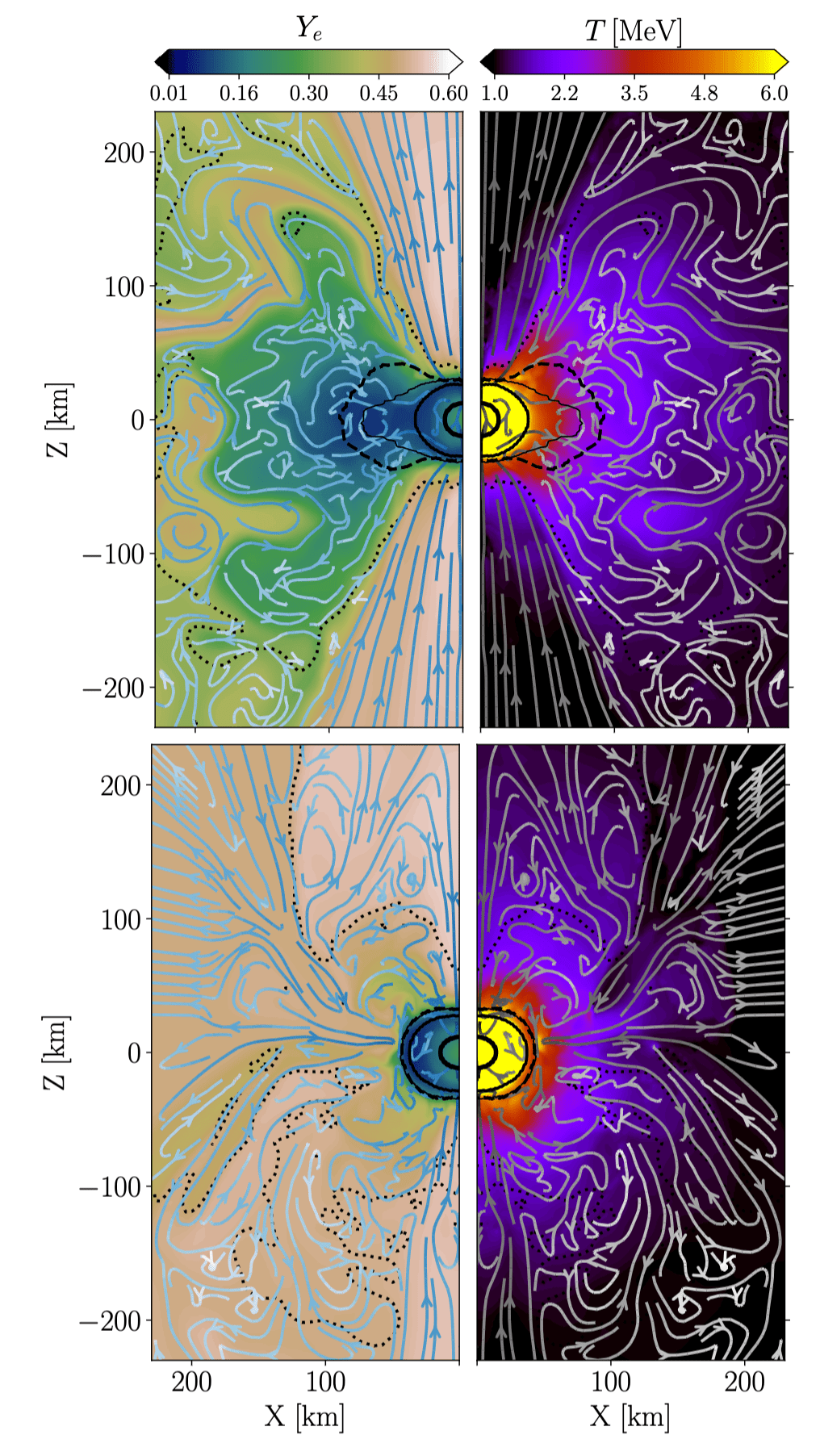}
\caption{Electron fraction (left column) and temperature (right) at $t\sim525$ ms p.b., for models \modelfe{1} (top row) and \modelfe{4} (bottom). 
The solid black lines represent density contours of $10^{11}$, $10^{12}$ and $10^{14}\,\rm{g\ cm}^{-3}$. 
The dashed line is the electron neutrino-sphere, while the dotted lines enclose the gravitationally unbound ejecta. 
The streamlines represent the poloidal component of the magnetic field.}
\label{fig:ye-T}
\end{figure}

Given the complexity of the dynamics of the collapse, it is rather difficult to unequivocally define the still forming PNS. 
The two most popular approaches that can be found in the literature identify the PNS surface with either the electron neutrino-sphere (defined as the locus of points where the total optical depth for electron neutrinos equals one) or a particular threshold $\rho_\tx{th}$ in mass density (ranging from $10^{11}$ to $10^{12}\,\rm{g\ cm}^{-3}$).
We opted for the latter strategy, by setting $\rho_\tx{th}=10^{11}\,\rm{g\ cm}^{-3}$, noticing that it provides a similar estimate for the PNS surface to the one using the neutrino-sphere while being overall less prone to fluctuations. 

\refig{fig:j-rho} and \refig{fig:ye-T} show some characteristic differences in the surroundings of the PNS between models \modelfe{1} (top panels) and \modelfe{4} (bottom panels) at a relatively late stage of the evolution ($\sim500\,\rm{ms}$ p.b.), which arise in general amongst models employing magnetic fields with different angular distributions. 
Models with low order multipoles tend to produce, within the duration of the numerical simulation, much more oblate PNSs and hence neutrino-spheres. However, the size of the PNS along the polar direction does not change accordingly.
A more oblate PNS is also surrounded by a region of gravitationally bound, denser and neutron-rich material that could, at later times, form a rotationally supported torus orbiting around the PNS (as we can see from the high specific angular momentum $j$ in the top left panel of \refig{fig:j-rho}). 
On the other hand, model \modelfe{4} does not show any of these features, in analogy with all the cases with higher values of $l$. 
In this case the PNS appears much more spherical, although it is still deformed at the equator by the fast rotation. 
There is also no sign of a rotationally supported structure surrounding the central object, with the neutron-rich material being well confined within the neutrino-sphere.  

\begin{figure}
\centering
\includegraphics[width=1.\columnwidth]{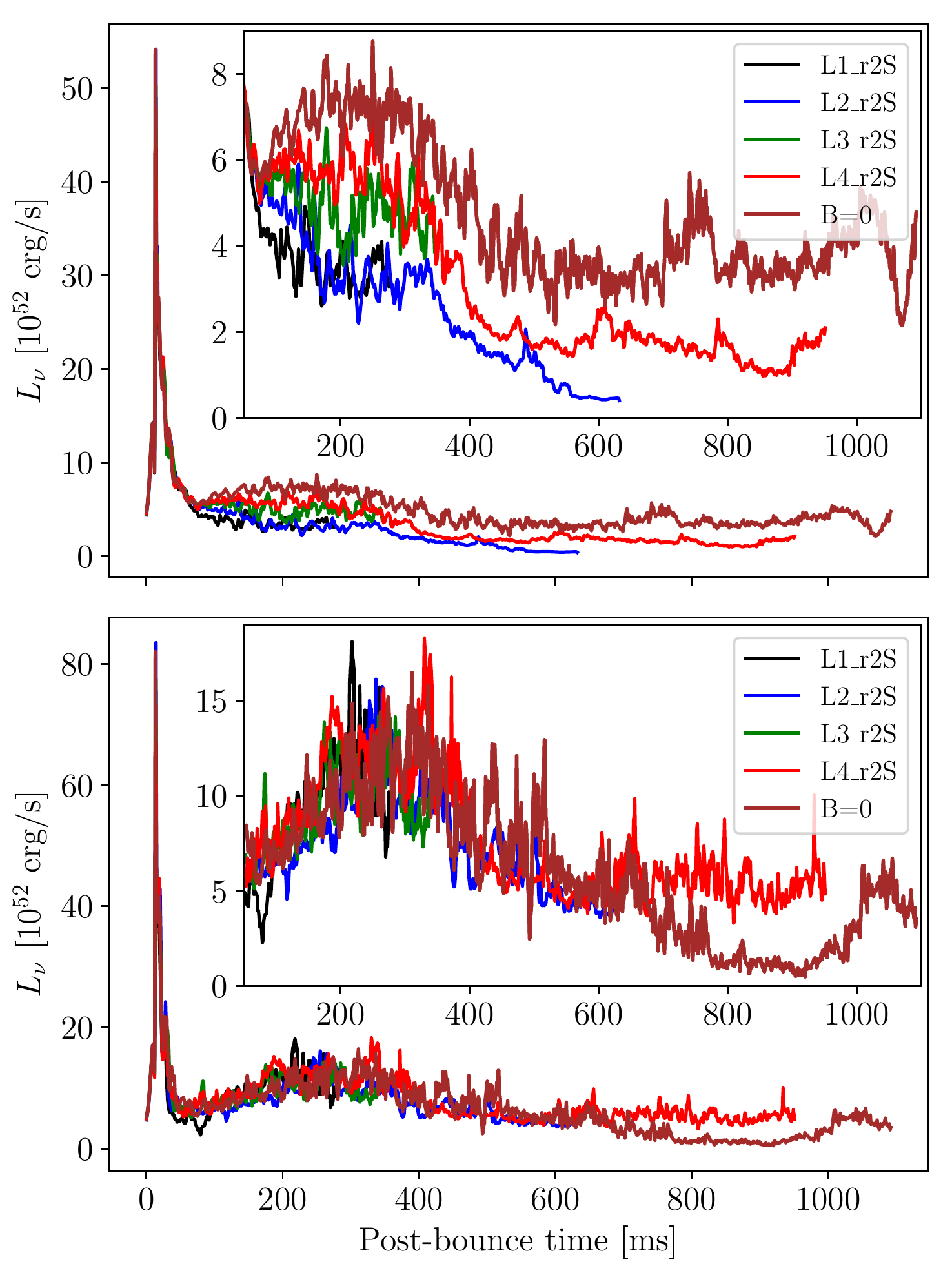}
\caption{Time evolution of the electron neutrino luminosity at a distance of $500\,\rm{km}$ along the equatorial plane (top) and in the north direction (bottom) for models \modelpns{\text{*}}.}
\label{fig:luminosity}
\end{figure}

Another consequence of the more oblate shape of the PNS for low multipolar configurations is a significantly lower temperature at the neutrino-sphere along the equatorial plane (see right panels in \refig{fig:ye-T}).
While for model \modelfe{1} we see a temperature of $\sim2.3\,\rm{MeV}$, model \modelfe{4} displays instead a value of $\sim3.9\,\rm{MeV}$.
This has as a direct consequence the emission of less energetic neutrinos in the equatorial plane, which deeply affects the related neutrino luminosity.
Along the equator (top panel of \refig{fig:luminosity}) models with higher order multipoles show a systematically larger luminosity, while along the symmetry axis (bottom panel) such a difference does not occur. 
Moreover, the hydrodynamic case presented the highest luminosity along the equatorial plane, since it is the model that produces the least oblate PNS amongst the ones we considered in the present work.

\begin{figure}
\centering
\includegraphics[width=0.95\columnwidth]{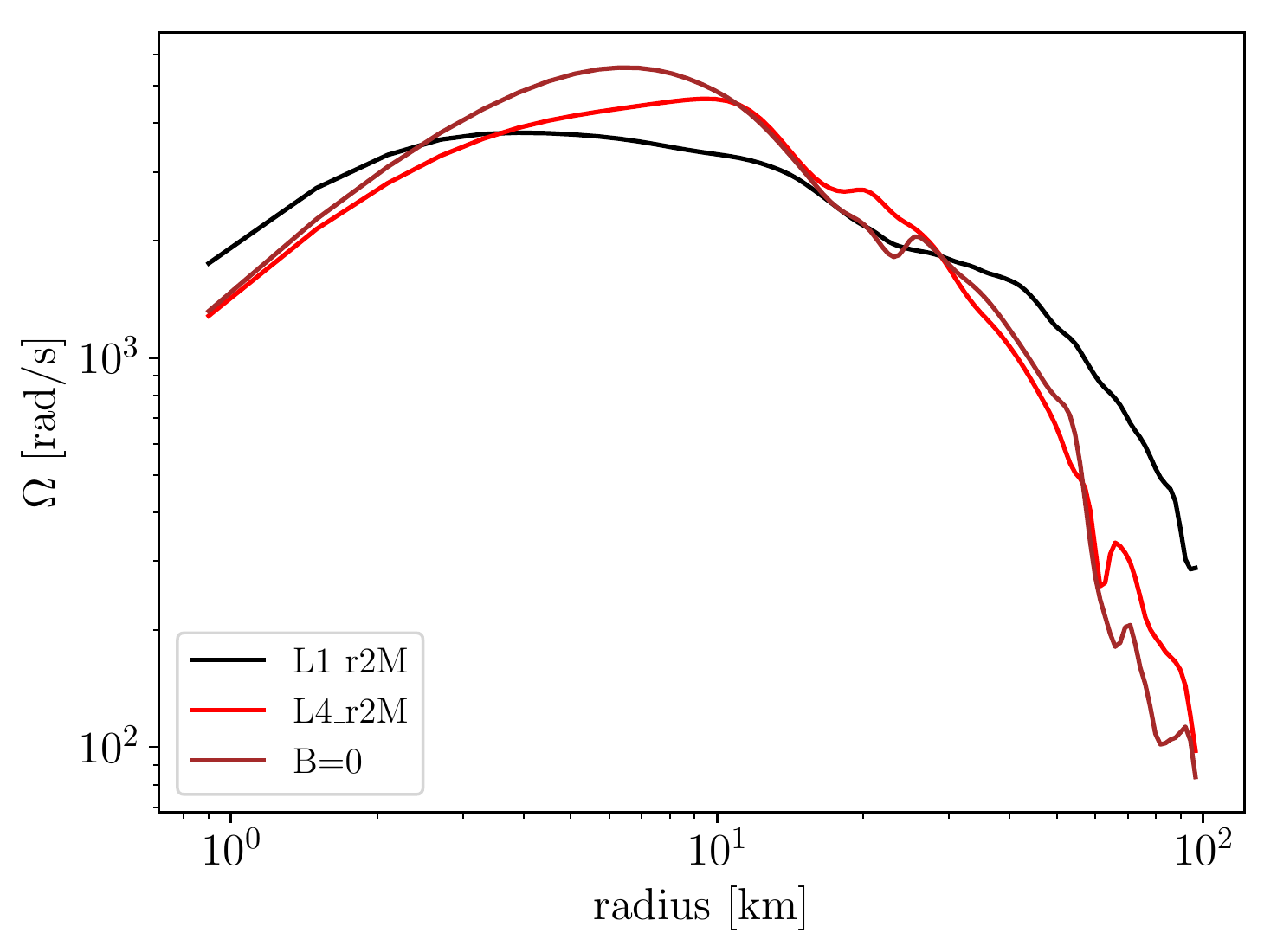}
\caption{Radial profile of the angular velocity $\Omega$ in the equator for models \modelfe{1} (black), \modelfe{4} (red) and the hydrodynamic model (brown) at $t\sim400\,\rm{ms}$ p.b..}
\label{fig:omega_equator}
\end{figure}

The torus-like structure surrounding the PNS in the presence of a strong dipolar field can be traced back to the radial profile of the angular velocity.
\refig{fig:omega_equator} shows that the region in proximity of the PNS rotates faster for models with lower initial value of $l$.
Moreover, the interior of the PNS exhibits a much shallower rotation profile approaching solid-body rotation, as opposed to the higher $l$ models and the hydrodynamic case which present an increase with radius of $\Omega$ up to a certain radius, followed by a steep decay.
The differences in the rotation profiles are due to a more effective angular momentum transport for lower order multipoles. 
This can be interpreted by the fact that the dipolar magnetic field is coherent over larger scales than the higher order multipoles, hence resulting in a more effective winding and subsequent transport of angular momentum. 
Note that, although the effect is weaker for higher order multipoles, these models also show at later times significant angular momentum transport resulting in a flattening of the equatorial profile of $\Omega$ if the magnetic field is stronger with $B_0>10^{11}$G.

\begin{figure}
\centering
\includegraphics[width=1.\columnwidth]{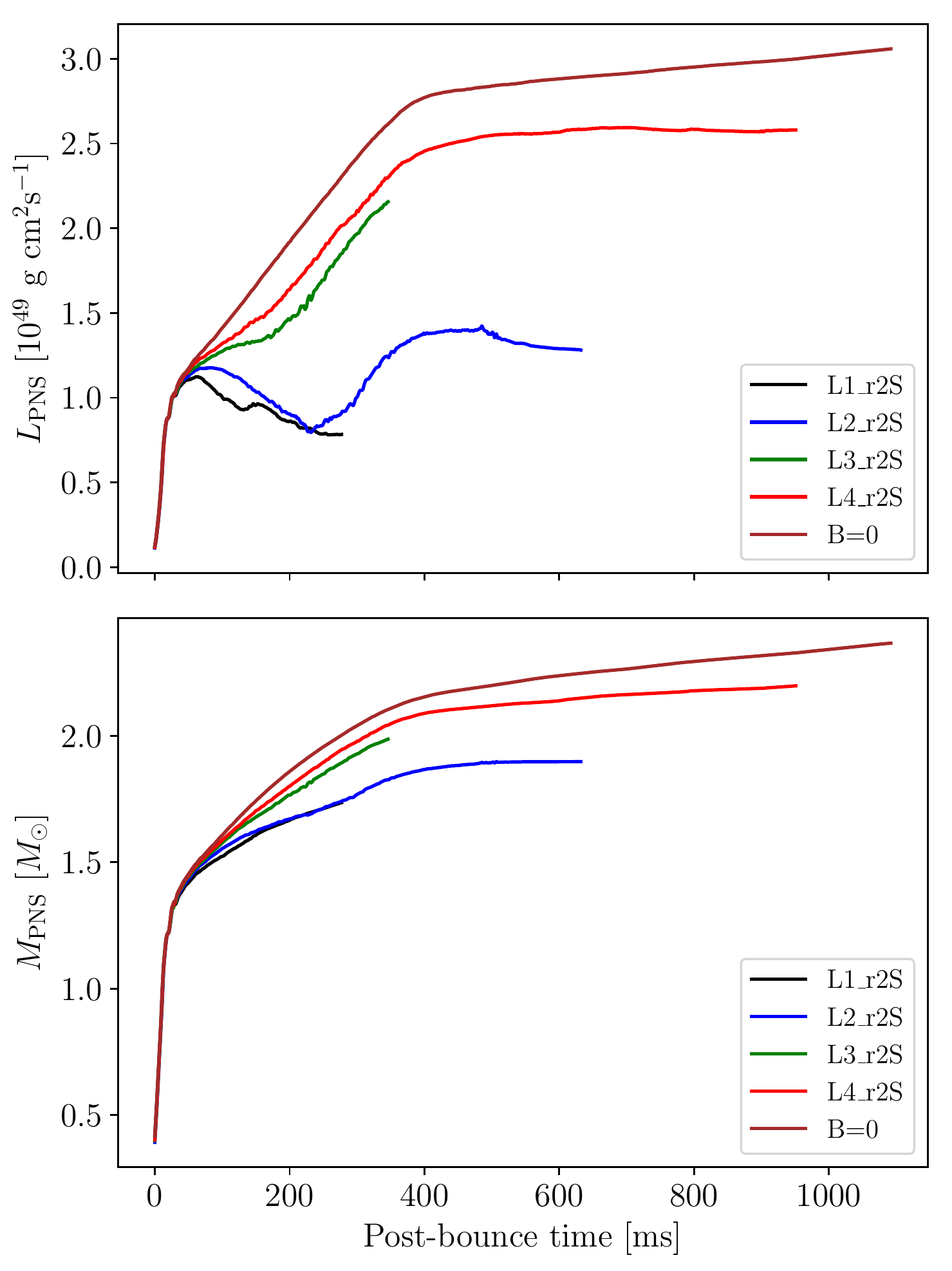}
\caption{Time evolution of the total angular momentum and mass of the PNS for models \modelpns{\text{*}}.}
\label{fig:pns_angmom}
\end{figure}

The evolution in time of the PNS total angular momentum $L_\tx{PNS}$ gives some more quantitative insights on the efficiency of rotational energy extraction from the central compact object in order to power up the polar outflows. 
The braking of the PNS rotational motion by the magnetic stresses appears to be quite sensitive to the topology of the magnetic field, as can be seen in the top  panel of \refig{fig:pns_angmom} by comparing the hydrodynamic model with models \modelpns{\text{*}}. 
In the non-magnetized case, the PNS angular momentum increases continuously due the accretion of rotating material on the PNS surface.
Indeed, even without magnetic fields the neutrino-driven explosion we produce in our hydrodynamic model is still rather asymmetric, with most of the ejecta expanding along the symmetry axis and an ongoing accretion flow at low latitudes.
The reason behind this persistent asymmetry is the combination of strong rotation, which induces an asymmetry between the poles and the equator \citep[e.g.][]{suwa2010}, and  the assumption of axisymmetry \citep{summa2016,bruenn2016,vartanyan2018,oconnor2018}, which naturally favours the development of oscillating modes along the symmetry axis. 
The change in slope occurring around $\sim400\,\rm{ms}$ p.b. is a consequence of the accretion of the interface between the iron core and the external convective shell, which is in general much less dense than the iron core and in the particular case of the progenitor 35OC presents a sudden decrease in specific angular momentum.
For sufficiently strong magnetic fields (i.e. $B_0>3\times10^{10}\,\rm{G}$) the PNS angular momentum for all multipolar configurations is always appreciably smaller than the hydrodynamic counterpart. 
In the case with a dipolar magnetic field, the initial growth of the PNS total angular momentum is halted at about $50\,\rm{ms}$ p.b., when it starts to decrease rather sharply. 
For larger values of $l$, however, this effect is significantly mitigated, as they show a faster restart of the angular momentum growth (the sooner the higher is $l$) that is associated with a decrease in the efficiency of the magnetic braking mechanism. 
The lower $L_\tx{PNS}$ can be directly linked to the larger explosion energy shown in \refig{fig:exp_energy}: a more efficient magnetic braking results in more rotational energy extracted from the PNS and hence a more energetic polar outflow.

The growth of the total mass of the PNS is highly correlated to that of the PNS angular momentum. 
Models with a less effective magnetic braking (either because of a weaker initial fields or a higher order of multipolar expansion) present an increasingly faster growth of the mass accreted onto the PNS, although they still lead to less massive PNS with respect to the case with no magnetic fields (provided that the initial field is sufficiently strong, as shown for the total angular momentum). 
In some cases (i.e. for models \model{1}{12}, \modelpns{2} and \modelfe{1}) the PNS mass appears to reach a plateau after a few hundred of ms p.b at no less than $\sim1.9$ $M_\odot$, suggesting that a collapse to BH could at least be significantly postponed, if not even prevented at all. 

\begin{figure}
\centering
\includegraphics[width=0.95\columnwidth]{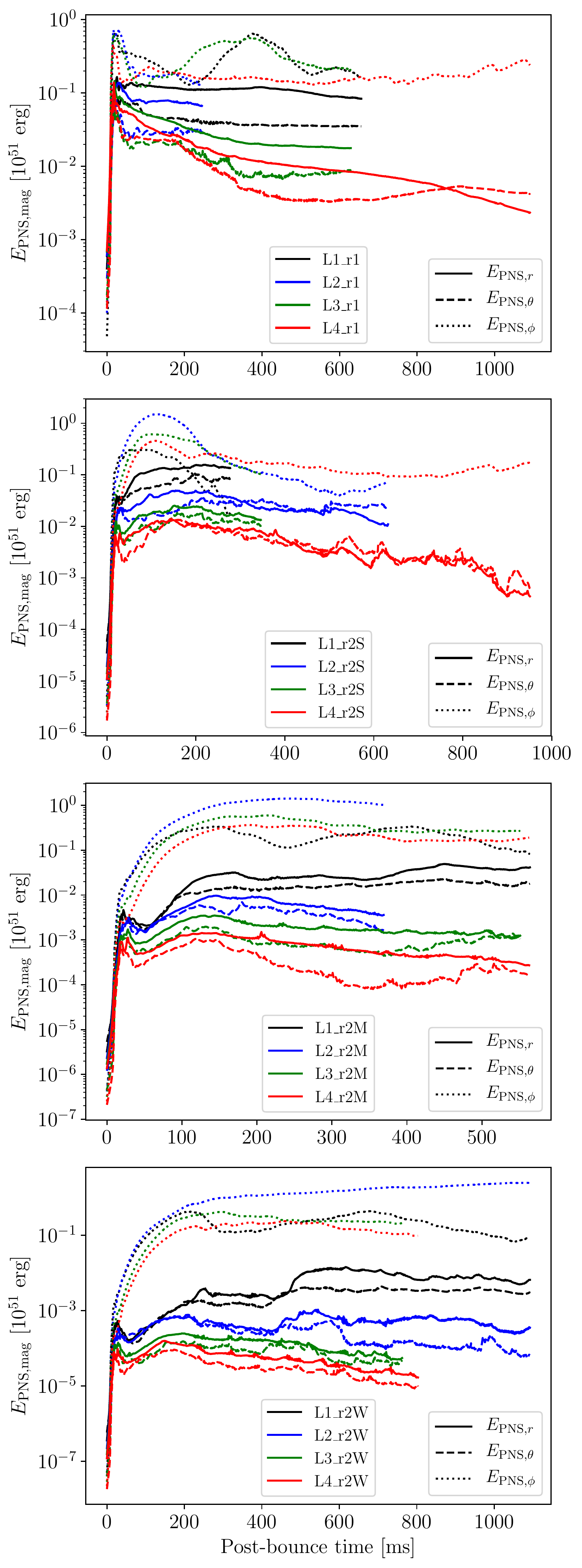}
\caption{Time evolution of the magnetic energy density integrated over the PNS volume. 
Solid, dashed and dotted lines represent the contribution respectively of the radial, polar and azimuthal component of the magnetic field. 
As in \refig{fig:radii}, the top panel refers to different multipolar configurations with a small value for $r_0$ and a strong field (models \model{\text{*}}{12}), while the other three show (from top to bottom) models with large $r_0$ and progressively weaker magnetic field strength.}
\label{fig:pns_emag}
\end{figure}

%%%%%%%%%%%%%%%%%%%%%%%%%%%%%%%%%%%%%%%%%%%%%%%%%%
%%%%%%%%%%%%%%%%%%%%%%%%%%%%%%%%%%%%%%%%%%%%%%%%%%
%%%%%%%%%%%%%%%%%%%%%%%%%%%%%%%%%%%%%%%%%%%%%%%%%%

\subsection{Evolution of the magnetic field}
We now analyse more in detail the dynamic evolution of the magnetic field during our simulations, focusing particularly on the PNS and its immediate surroundings. 
\refig{fig:pns_emag} shows the contributions to the magnetic energy stored within the PNS due to the three spatial components of the field. 
For models \model{\text{*}}{b12} (top panel), which start at bounce with approximately the same content in magnetic energy, the contributions of the poloidal components quickly diverge for different multipolar orders. 
The dipole model exhibits rather constant radial and polar components, while the higher multipoles show an increasingly faster decay of both, leading to a spread of more than an order of magnitude. 
On the other hand, the toroidal components of the magnetic energy (dotted lines) are much closer to each other.

For all other models with larger value of $r_0$ we observe a transient growth of the poloidal magnetic field right after bounce, whose duration spans approximately  $\sim50-150\,\rm{ms}$, depending on the particular setup. 
This is due to the conservation of magnetic flux during the accretion of still highly magnetised material (which is missing in the case of models \model{\text{*}}{12}) and further contraction of the PNS. 
On the other hand, the toroidal component of the magnetic field undergoes a steady growth due to the \emph{$\Omega$-effect} via a a term in the induction equation of the form $\sim r\sin\theta B\nabla\Omega$, where $B$ is the magnetic field component parallel to $\bm{\nabla}\Omega$.
Despite these poloidal components being weaker than in models \model{\text{*}}{b12}, the toroidal field generated by the winding of field lines is comparable.
For most models the toroidal component  appears to saturate at a similar value, regardless of the specific initialisation of the magnetic field.
Only in the case of a quadrupolar field ($l=2$) and sufficiently weak fields ($B_0\lesssim10^{11}$ G) the toroidal component is considerably larger than all the other configurations. 
This effect could result from a more efficient winding because of a non-vanishing radial magnetic field in the equatorial region (where $\bm{\nabla}\Omega$ is mostly radial) due to the even value of $l$ in combination with a distribution of the field on larger angular scales than other even-$l$ (i.e. $l=4$), and hence less dissipation. 

As in models \model{\text{*}}{b12}, dissipation of the poloidal magnetic energy occurs faster for higher multipolar configurations, a trend that shows up for stronger initial magnetic fields as well (second panel from the top). 
By contrast, we do not observe an appreciable dissipation of the poloidal magnetic energy in simulations with either a weaker magnetic field or a larger scale distribution. 
A more detailed investigation showed that most of this dissipation occurs in the deep inner part of the PNS. 
This part of the numerical domain has a grid coarsened in the lateral direction (so as to keep approximately square numerical cells), which could be a possible source of intrinsic numerical dissipation \citep[see][for a thorough investigation of the numerical diffusion in this code]{rembiasz2017}. 
Part of the dissipation of the magnetic field may also be attributed to a turbulent resistivity since convective motions are present in this region. 
In both cases, the faster dissipation of the higher multipoles can be explained by the smaller scale of the field since the dissipation rate scales like $k^2\eta$, with the wavenumber $k^2 \sim l(l+1)/r^2$ increasing with $l$. 
However, since this dissipation is limited to the deep interior of the PNS, we do not expect it to affect significantly the dynamics of the explosion, which is driven by the surface layers of the PNS. 
The fact that we use a fifth-order spatial reconstruction algorithm and the lateral resolution in the surface layer should be sufficient to minimize numerical dissipation, as even a $l=4$ field is described by 64 cells per wavelength.

\begin{figure}
\centering
\includegraphics[width=\columnwidth,]{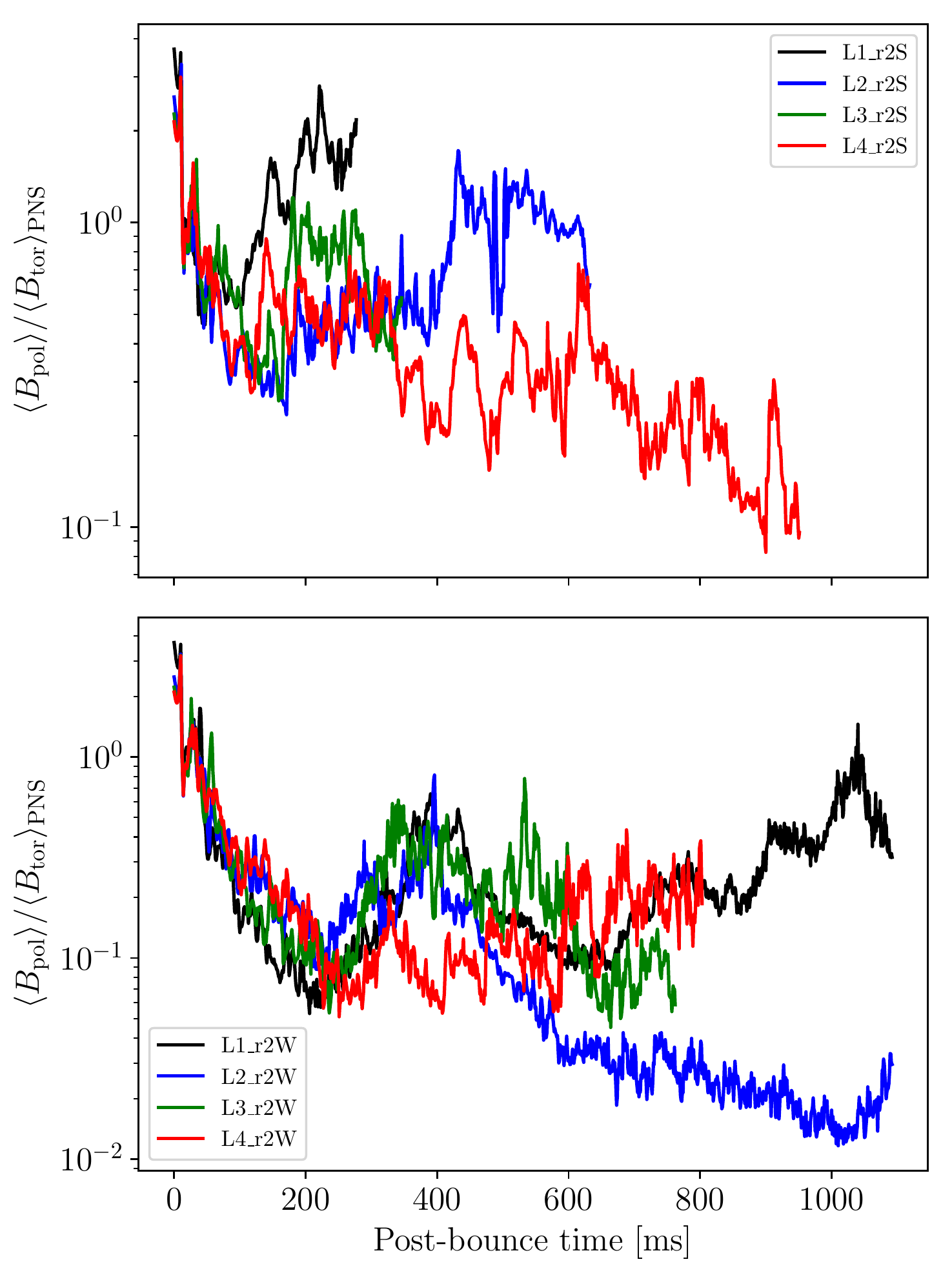}
\caption{Time evolution of the ratio of the averaged poloidal and toroidal components of the magnetic field over the PNS surface for models \modelpns{\text{*}} (top panel) and \modelweak{\text{*}} (bottom one).}
\label{fig:pns_poltor}
\end{figure}

{\refig{fig:pns_poltor} shows the ratio of the averaged poloidal and toroidal magnetic fields, where the average is performed over the PNS surface.
During the stalling of the shock it undergoes a systematic decrease, whose duration goes from being almost null for model \model{1}{12} to about $200\,\rm{ms}$ for the models with very weak initial field. 
This is due to the continuous increase of the toroidal field produced by the $\Omega$-effect, which is the more effective the longer the magnetised layers accreting onto the PNS stay in the highly differentially rotating region beyond the shock. 
Afterwards, the ratio increases in coincidence with the start of shock expansion, as a consequence of the expulsion of gas magnetised with the strong toroidal component produced by the winding of the magnetic field lines. 

To better understand how the initial topology of the field changes throughout our simulations, we compute the angular spectra of the radial component of the magnetic field as
\be
 B^r_l(r,t)=\int B^r(r,\theta,t)Y_l^0(\theta)\tx{d}\theta,
\ee
where $Y_{l0}$ is the spherical harmonic of order $l$ in axisymmetry.
We then focus again on the dynamics of the field near the PNS surface by averaging these quantity over the radius in a small region surrounding the surface of the PNS, i.e.
\be\label{eq:spectra}
\tilde{B}^r_l(t) = \frac{\int_{r_\tx{min}}^{r_\tx{max}}B^r_l(r,t)\tx{d}r}{\int_{r_\tx{min}}^{r_\tx{max}}\tx{d}r},
\ee
where $r_\tx{min}=0.95R_\tx{pns}$ and $r_\tx{max}=1.05R_\tx{pns}$. 

In \refig{fig:spectra_Br} we report spectrograms (time vs. multipole order $l$) computed for models \modelfe{\text{*}}.
At each given time we normalised each spectrum by the total power among different modes, in order to appreciate the shift in relative importance among different multipoles with time.
We first notice that the power in the initially dominant multipolar component (i.e. the red spot at the bottom of each diagram) is quickly spread among smaller scales.
Harmonics with the same parity as the initial field get excited, although we also see a few modes with opposite parity undergoing some sporadic growth. 
The initial multipole is, in general, always overtaken by higher order modes. 
Only in the dipolar case the $l=1$ mode remains comparable in power to the strongest one (that is the $l=3$ mode), showing a much weaker composition on the smaller scales than the other models reported in the figure.
For models with $l>1$ we also see the emergence of a radial component on larger scales, although it fails to become dominant within the duration of our simulations. 
In addition, the spectra of the models with more complex topology peak around $l\sim10$, showing a strong excitation of modes on smaller angular scales and a positive spectral slope at low $l$.  
These results suggest that the radial field at the PNS surface is unlikely to retain a dominant component at the largest angular scales, unless it starts with a dominant dipolar component. 

\begin{figure}
\centering
\includegraphics[width=\columnwidth]{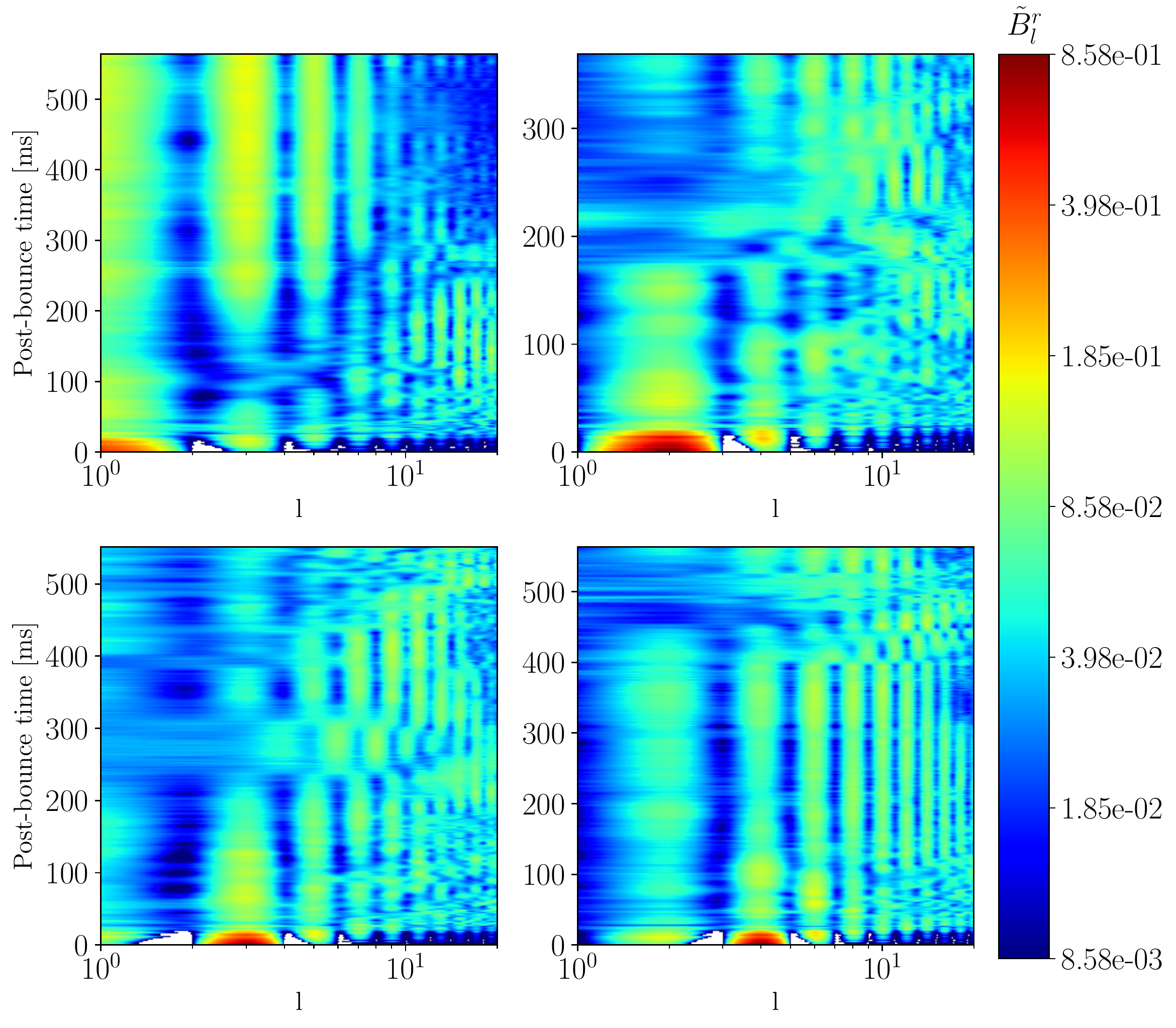}
\includegraphics[width=\columnwidth]{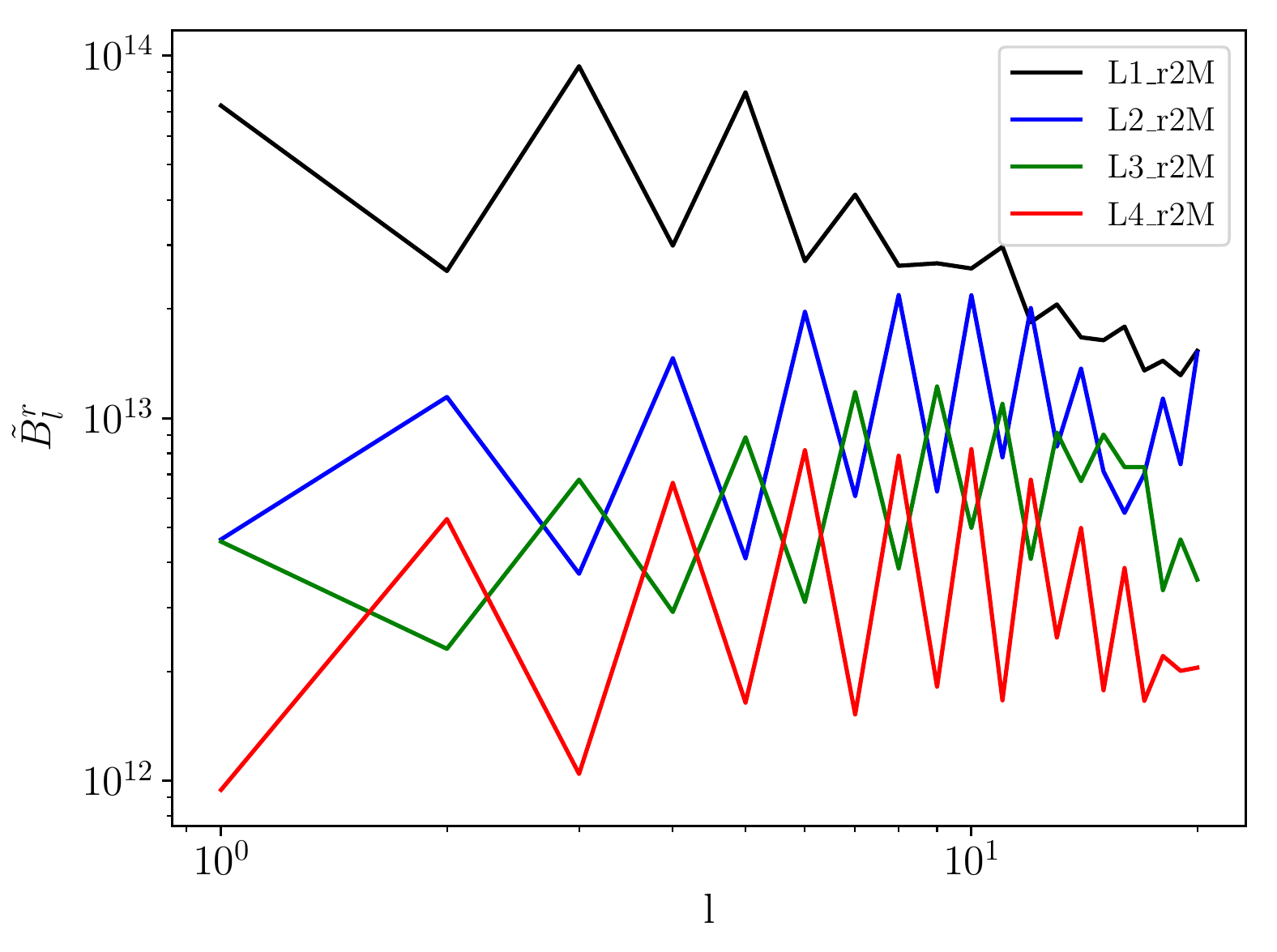}
\caption{Harmonic decomposition of the radial component of the magnetic field at the PNS surface for models \modelfe{\text{*}} as defined in \refeq{eq:spectra} (top panel) and spectra taken at $t=358\,\rm{ms}$ p.b. (bottom panel).}
\label{fig:spectra_Br}
\end{figure}

%%%%%%%%%%%%%%%%%%%%%%%%%%%%%%%%%%%%%%%%%%%%%%%%%%
%%%%%%%%%%%%%%%%%%%%%%%%%%%%%%%%%%%%%%%%%%%%%%%%%%
%%%%%%%%%%%%%%%%%%%%%%%%%%%%%%%%%%%%%%%%%%%%%%%%%%
%%%%%%%%%%%%%%%%%%%%%%%%%%%%%%%%%%%%%%%%%%%%%%%%%%
%%%%%%%%%%%%%%%%%%%%%%%%%%%%%%%%%%%%%%%%%%%%%%%%%%
%%%%%%%%%%%%%%%%%%%%%%%%%%%%%%%%%%%%%%%%%%%%%%%%%%
%%%%%%%%%%%%%%%%%%%%%%%%%%%%%%%%%%%%%%%%%%%%%%%%%%
%%%%%%%%%%%%%%%%%%%%%%%%%%%%%%%%%%%%%%%%%%%%%%%%%%
%%%%%%%%%%%%%%%%%%%%%%%%%%%%%%%%%%%%%%%%%%%%%%%%%%
%%%%%%%%%%%%%%%%%%%%%%%%%%%%%%%%%%%%%%%%%%%%%%%%%%

\section{Conclusions}
\label{sec:conclusions}

We presented a set of axisymmetric special relativistic MHD simulations of core-collapse supernov\ae\ with the goal of assessing the impact of the magnetic field's angular distribution and radial extent on the explosion dynamics and the formation of the central PNS. 

Models initialised with increasingly high order multipolar configurations generally produce weaker explosions, slower expanding shocks and less collimated outflows.
Even when the multipolar configuration matches the magnetic energy content of the dipolar one, a similar difference is present in both the shock expansion and the explosion energy.
The contribution of magnetic pressure to the onset of the explosion decreases if the initial field has a spatial distribution on smaller angular scales. 
We interpret this effect as due to the higher multipoles connecting a smaller fraction of the PNS surroundings to the polar region, hence leading to a smaller degree of magnetisation in proximity of the rotational axis.

In addition to this dependence on the angular structure of the magnetic field, we observed an important impact of its radial distribution. 
Larger radial extent of the initial magnetic field allows for a more effective compression and hence amplification of the field. 
By contrast, in models \model{\text{*}}{12} the steeper radial decay of the magnetic field for higher multipolar orders produces qualitatively different scenarios.
If the highly magnetised central region of the progenitor is buried below less magnetised material, the PNS surface has a too weak magnetic field to produce a strong explosion.

The PNS can exhibit quite different features amongst the various models: the lower the multipolar expansion order $l$, the more oblate and elongated along the equatorial plane is the PNS. 
Some dipolar models show the early stages of formation of a torus-like structure with clear signs of rotational support but also a neutron rich composition.
This suggests that at later times this structure could result into an accretion torus. 
The more oblate shape has also the consequence of significantly decreasing the temperature at the neutrino-sphere, and thus the energy and luminosity of the neutrinos emitted along the equator. 
Once again, models with increasing multipolar order tend to approach the limit of a hydrodynamic explosion, which presents the highest neutrino luminosity along the equatorial plane.

The topology of the magnetic field has a direct impact on the growth of the PNS mass and spin. 
Larger scale configurations produce less massive PNS with slower rotation, due to a more effective magnetic braking. 
This dichotomy is fully consistent with the impact of magnetic topology on the explosion dynamics: an enhanced efficiency in extracting the rotational energy from the PNS (which occurs for lower multipoles) leads naturally to more energetic explosions and hence faster expanding shocks. 
Increasing the multipole order leads, on the contrary, to more massive and faster rotating central objects, with the tendency of approaching the limiting hydrodynamic case.
A different angular distribution of the magnetic field can, therefore, have an impact on the delay (and possibly the overall prevention) of the PNS collapse to a black hole.  

The radial component of the magnetic field at the surface of the PNS stops being dominated by the strongest mode in the initial multipolar expansion within a few hundred ms, as we observe a broadening of the harmonic spectrum to both smaller and (in some cases) larger scales. 
The spectrum appears to peak at larger scales ($l\sim3$) in the dipolar case with respect to the higher multipolar models (whose spectra peaks around $l\sim10$). 
Once again, we find a much more striking difference between the dipolar case and models with $l>1$ than between different multipolar configurations.

Despite being much less impactful than a magnetic dipole, higher multipolar configurations can still strongly affect the explosion dynamics and PNS formation.
While our results disfavour the scenario of complex magnetic field configurations leading to very energetic explosions, they allow for the possibility of magnetically driven explosions starting from higher multipolar fields. 
An interesting feature of SNRs associated with known magnetars is the lack of particular asymmetries or hints of polar ejection \citep{kaspi2017},
although the sample of such SNRs is limited to about 10 sources \citep{ferrand2012,beniamini2019}. 
Our results show that higher multipoles tend to deliver much less collimated polar outflows, which could reconcile the SNR observations with magnetorotational explosions by the action of magnetic fields with complex topologies (at least during the onset of the explosion). 

Another important feature of the higher multipolar configuration is a general lack of continuous growth of either the magnetisation or the baryon loading parameter in the outflow, due to a decrease in the magnetic and kinetic energy fluxes. 
Only in the case of a dipolar field we see a systematic growth of both $\sigma$ and $\eta$. 
It would be interesting to investigate the consequences that these trends in the evolution of the energy fluxes in the polar outflow might have on the launching of a magnetar-driven relativistic outflow.
A central protomagnetar could in fact be the central engine powering up GRB-like events \citep{metzger2011,metzger2018}, provided that the luminosity in the outflow reaches a critical value that allows the jet to break through the SN ejecta \citep{aloy2018}.
However, a longer evolution after core bounce is necessary to fully settle the fiducial values of the magnetisation by the time in which the central PNS becomes an active protomagnetar central engine.
We therefore leave this aspect to be investigated in future work.

Our results on the dependence on the angular distribution of the magnetic field stem from the same initial stellar evolution model. 
We have pointed out in the introduction that there is lot of room for setting up the magnetic field topology of the pre-collapse model. 
However, the fact that the outcome of stellar core collapse may vary so much changing the topology of the magnetic field claims for improved, genuinely three-dimensional stellar evolution models including magnetic fields. 
Once such models are available, more realistic magnetic field configurations may be used, hence making our models more predictive.

If on the other hand, the strong initial magnetic field is thought of as a proxy for a PNS dynamo, our results highlight the need for a better understanding of this process and particularly of the magnetic field topology. 
Recent numerical simulations of convective dynamos \citep{raynaud2019} and MRI \citep{reboul-salze2019} within the PNS showed the generation of complex magnetic configurations in which the dipolar component is not dominant. 
The fact that higher order magnetic configurations can still lead to magnetically driven explosions is an important result in this context. It shows that the magnetic field impact cannot be summarised only by its dipolar component and highlights the need to know the full magnetic field structure.

An important caveat to the present work is represented of course by the fact that all our results were obtained assuming axisymmetry. 
A priori, we could expect a quantitative change in our findings once we allow the system to evolve along the azimuthal direction as well.
It is still a matter of debate, however, whether the inclusion of non-axisymmetric dynamics should qualitatively impact the onset of the explosion. 
While \cite{mosta2014} find that the development of the kink instability in the outflow can prevent a successful expansion of the shock along the polar axis, \cite{obergaulinger2019,aloy2019b} find that a powerful magnetorotational explosion can still be produced in a three-dimensional domain. 
More recently, the impact of a misaligned magnetic dipole on the explosion has been investigated by \cite{halevi2018}, but it remains still not clear how a multipolar field would evolve once the axisymmetric condition is relaxed. For this reason we plan on studying this aspect in forthcoming work. 

%%%%%%%%%%%%%%%%%%%%%%%%%%%%%%%%%%%%%%%%%%%%%%%%%%
%%%%%%%%%%%%%%%%%%%%%%%%%%%%%%%%%%%%%%%%%%%%%%%%%%
%%%%%%%%%%%%%%%%%%%%%%%%%%%%%%%%%%%%%%%%%%%%%%%%%%
%%%%%%%%%%%%%%%%%%%%%%%%%%%%%%%%%%%%%%%%%%%%%%%%%%
%%%%%%%%%%%%%%%%%%%%%%%%%%%%%%%%%%%%%%%%%%%%%%%%%%
%%%%%%%%%%%%%%%%%%%%%%%%%%%%%%%%%%%%%%%%%%%%%%%%%%
%%%%%%%%%%%%%%%%%%%%%%%%%%%%%%%%%%%%%%%%%%%%%%%%%%
%%%%%%%%%%%%%%%%%%%%%%%%%%%%%%%%%%%%%%%%%%%%%%%%%%
%%%%%%%%%%%%%%%%%%%%%%%%%%%%%%%%%%%%%%%%%%%%%%%%%%
%%%%%%%%%%%%%%%%%%%%%%%%%%%%%%%%%%%%%%%%%%%%%%%%%%

\section*{Acknowledgements}

MB and JG acknowledge support from the European Research Council (ERC starting grant no. 715368 -- MagBURST) and from the \emph{Tr\`es Grand Centre de calcul du CEA} (TGCC) and GENCI for providing computational time on the machines IRENE and OCCIGEN (allocation A0050410317). 
MO acknowledges support from the European Research Council under grant EUROPIUM-667912, and from the Deutsche Forschungsgemeinschaft through Sonderforschungsbereich SFB 1245 "Nuclei: From fundamental interactions to structure and stars".
MO, PCD and MAA acknowledge the support through the grants AYA2015-66899-C1-1-P and PROMETEOII-2014-069 of the Spanish Ministry of Economy and Competitiveness (MINECO) and of the Generalitat Valenciana, respectively. JG and MAA acknowledge the support from the PHAROS COST Action CA16214, while MAA also acknowledges the GWverse COST Action CA16104.

%%%%%%%%%%%%%%%%%%%%%%%%%%%%%%%%%%%%%%%%%%%%%%%%%%



\bibliographystyle{mnras}
\bibliography{references}

%%%%%%%%%%%%%%%%% APPENDICES %%%%%%%%%%%%%%%%%%%%%

\appendix

%%%%%%%%%%%%%%%%%%%%%%%%%%%%%%%%%%%%%%%%%%%%%%%%%%

% Don't change these lines
\bsp	% typesetting comment
\label{lastpage}
\end{document}